\documentclass[a4paper,11pt]{article}
\usepackage{jheppub} 
\usepackage{lineno}
\usepackage{amsmath,amssymb,amsthm}
\usepackage{color}
\usepackage{amsfonts}
\usepackage{tikz-cd}
\usepackage{cancel}
\usepackage{arydshln}
\usepackage{simpler-wick}
\usepackage{booktabs}
\usepackage{slashed}
\usepackage{feynmp}
\usepackage{xcolor}

\DeclareMathOperator{\tr}{\textrm{tr}}
\DeclareMathOperator{\Tr}{\textrm{Tr}}
\DeclareMathOperator{\Str}{\textrm{Str}}

\arxivnumber{} 

\title{Beta-deformation in Twistor-String Theory}

\author{Eggon Viana}
\affiliation{{\it  ICTP South American Institute for Fundamental Research\\
Instituto de F\'{i}sica Te\'{o}rica, UNESP-Universidade Estadual Paulista\\
R. Dr. Bento T. Ferraz 271, Bl. II, S\~{a}o Paulo 01140-070, SP, Brazil\\

Instituto Gallego de Física de Altas Energías (IGFAE),\\
Departamento de Física de Partículas,\\
Universidad de Santiago de Compostela
}}

\emailAdd{eggon.viana@unesp.br}

\abstract{In this work, we investigate how the marginal beta deformation of the \(\mathcal{N}=4\) super-Yang-Mills theory manifests within the context of the topological B-model in the twistor space \(\mathbb{CP}^{3|4}\). We begin by identifying the beta deformation as states living in a specific irreducible representation of the superconformal algebra. Then, we compute the ghost number two elements of the BRST cohomology of the topological model. A gauge-fixing procedure is applied to these states, allowing us to identify the elements living in the irreducible representation that characterizes the beta deformation. Based on this identification, we proceed to write the deformed topological action, and the corresponding deformed BRST operator.}

\begin{document}
\maketitle
\flushbottom

\section{Introduction}

The beta deformation was initially described in the context of $\mathcal{N}=4$ Super-Yang-Mills theory by Leigh and Strassler in \cite{Leigh-Strassler}. This deformation plays a significant role in AdS/CFT \cite{Maldacena:1997,Witten:1998}, as it opens new possibilities to investigate concrete examples of the duality. From the perspective of the gauge/gravity duality, this deformation has a corresponding supergravity description, which was introduced by Maldacena and Lunin in \cite{Maldacena-Lunin}. Moreover, the corresponding string description of this deformation has been extensively studied in \cite{andrei-rivelles, Benitez, Viana}, using the pure spinor description of the superstring.

Still in the context of gauge/gravity duality, it is well known that there is a certain equivalence between $\mathcal{N}=4$ Super-Yang-Mills theory and the twistor string given by the B-model on $\mathbb{CP}^{3|4}$, which contains conformal supergravity \cite{confgra}. Indeed, the perturbative expansion of $\mathcal{N}=4$ SYM theory is equivalent to the D-instanton expansion of the B-model on $\mathbb{CP}^{3|4}$. Moreover, the physical spectrum of this model can be mapped, via Penrose transform, to the spectrum of $\mathcal{N}=4$ SYM \cite{twistor}. Therefore, it is quite natural to ask for the description of the beta deformation in the context of the twistor string \cite{Kulaxizi:2004,Gao}.

The physical states of the B-model on $\mathbb{CP}^{3|4}$ are defined as elements of the sheaf cohomology group in projective space \cite{twistor}. Moreover, the usual twistor space wavefunctions are only regular on an open subset of the projective space. For instance, the conformal supergravity spectrum in the twistor string follows from computing this type of cohomology \cite{confgra}. Therefore, as elements of a cohomology defined on a non-compact space, these states live in an infinite-dimensional representation of the superconformal algebra.

On the other hand, we know that the $\beta$-deformation states of string theory are described by vertex operators that live in a finite-dimensional representation of the superconformal algebra \cite{andrei-rivelles, Mikhailov:2009_vertex}. Therefore, the appropriate description of the twistor string states corresponding to the beta deformation is given in the context of cohomology groups defined on a compact space, as these cohomology groups are finite. In this paper, we will show this by identifying the beta deformation as elements of the cohomology defined on the entire projective space $\mathbb{CP}^{3|4}$, which is compact.

\subsection{Beta deformation in string theory}

From the point of view of string theory, the beta deformation can be understood as physical states that deform a string action defined on a given background. For instance, we can define the beta deformation in the context of the pure spinor superstring in $AdS_5 \times S^5$ \cite{andrei-rivelles}. Roughly speaking, the pure spinor formalism \cite{Berkovits_2001} is a manifestly supersymmetric description of the superstring, which contains a BRST symmetry. The BRST charge $Q$ has the property of being nilpotent, meaning $Q^2=0$. Physical states can be identified as elements in the BRST cohomology; that is, operators $V$ that are invariant under $Q$ and are not BRST variation of any other operator:
\begin{equation}
QV=0,\ \ \ \ \ V\ne Q\Lambda.
\end{equation}

Additionally, the superstring in \(AdS_5 \times S^5\) has a global symmetry realized through the Lie algebra of \(PSU(2,2|4)\). The conserved currents arising from this global symmetry are denoted \(j_a\), where \(a\) labels the Lie algebra \(\mathfrak{psu}(2,2|4)\). An important feature of the current \(j_a\) is that it satisfies the following equation under BRST variation:
\begin{equation}\label{Qj}
Qj_a = d\Lambda_a,
\end{equation}
for a given expression \(\Lambda_a\). In this context, the beta deformation can be introduced as physical states constructed from the conserved currents \(j_a\). These physical states were introduced in \cite{andrei-rivelles,Mikhailov:2009_vertex} and are defined as: 
\begin{equation}
U = B^{ab}j_a\wedge j_b,
\end{equation}

Using the relation (\ref{Qj}) above, we can see that $Q U = d(B^{ab} \Lambda_a j_b)$, and therefore it is BRST-invariant if we integrate it over the worldsheet $\Sigma$. The element $\int_\Sigma U$ is thus a physical state and deforms the action in the following way:  
\begin{equation}\label{S0}  
S = S_0 + \int_\Sigma U,  
\end{equation}
where $S_0$ is the original action in $AdS$. It is possible to see that, for specific values of $B^{ab}$, the deformation (\ref{S0}) reproduces the Maldacena-Lunin solution, which is the gravity dual of the beta deformation \cite{Viana}.

One may now ask whether this approach can be reproduced in other contexts. The aim of the present paper is to provide an affirmative answer for the case of a topological string described by the B-model on the projective space $\mathbb{CP}^{3|4}$. The main idea is that, as in the pure spinor string, this topological model carries the important features that are used to describe the beta deformation. Specifically:  
\begin{itemize}
    \item It has a BRST operator, which gives rise to the physical states;  
    \item It has a global symmetry under the superconformal algebra $\mathfrak{psl}(4|4)$, which results in conserved currents $j_a$;
    \item It has a well-defined composite $b$-ghost, which plays an important role in the gauge fixing of the physical states.  
\end{itemize}

The problem of investigating marginal deformations of $\mathcal{N}=4$ Super-Yang-Mills from twistor strings was first approached in \cite{Kulaxizi:2004} and later refined in \cite{Gao}. There, the beta deformation is understood as a correction to the holomorphic Chern–Simons defined on the projective space, which is the string field theory of the twistor string. In particular, at linear order in the deformation parameter, the effect of the deformation on the MHV amplitudes yields the expected results. In this paper, we will examine a similar problem, but now from the perspective of the worldsheet theory, which has $N=2$ supersymmetry. Moreover, we will provide a broader investigation through the possible deformations. Indeed, our description encompasses a full multiplet of deformations, which includes the Leigh-Strassler (LS) deformation studied in \cite{Kulaxizi:2004}. However, we are more general and our multiplets also includes other deformations. The first example is the $\gamma_i$-deformation, which is a 3-parameter version of the LS deformation \cite{Fokken:gammai}; the second example is the non-conformal deformation that gives rise to the non-commutative Yang-Mills theories \cite{vanTongeren,twist-noncommutative}. What these theories have in common is that they all preserves the integrability structure, and are denoted by integrable deformations \cite{Benoit:letter,q-deformation,Klimcik:YBT,Klimcik:YB,Hoare:notes,borsato-jj}. Although we don't approach this problem here, in the future we also expect to be able to cover other types of integrable deformations, such as the $\lambda$-deformation.

\subsection{Plan of the paper}

In Section 2, we review the definition of the beta deformation as states living in an irreducible representation of a symmetry algebra. In Section 3, we provide an overview of the topological B-model defined on the projective space $\mathbb{CP}^{3|4}$. In this section, the action, the BRST operator, and the $b$-ghost are introduced. In Section 4, we compute the cohomology of the BRST operator introduced in Section 2, with particular attention given to the ghost number 1 and 2 operators. In Section 5, we study the ghost numbers 1 and 2 in local coordinates, which is crucial for achieving the physical interpretation of these operators. Specifically, we closely investigate the ghost number 2 operators, which, in local coordinates, are manifestly understood as the states of the beta deformation. In Section 6, we offer general considerations on how the ghost number 2 operators deform the action and the BRST operator. In this section, we show that the topological action is deformed by a current-current term. Finally, in section 7 we make a discussion about further application of our results to the context of integrable deformations.

\section{Review of beta deformation}\label{betadeformation}

From the AdS/CFT correspondence, the \(\mathcal{N}=4\) SYM theory is dual to a type IIB superstring theory defined on the \(AdS_5 \times S^5\) space. The string dual of the beta deformation was studied in \cite{andrei-rivelles, Benitez} using the pure spinor formalism. The low-energy limit of this string description \cite{Viana} correctly reproduces the supergravity solution of Maldacena and Lunin \cite{Maldacena-Lunin}. In this section, we will properly define the beta deformation and explain how we plan to extend this description in the context of the topological B-model.

One can define the beta deformation in the context of string theory on the \(AdS_5 \times S^5\) space as states living in an irreducible representation of the \(\mathfrak{psu}(2,2|4)\) algebra. This irreducible representation was described in \cite{andrei-rivelles} as a subset of the antisymmetric product of two adjoint representations:  
\begin{equation}
\text{$\beta$-deformation} \subset \mathfrak{g} \wedge \mathfrak{g},
\end{equation}
where \(\mathfrak{g}\) denotes the adjoint representation of the \(\mathfrak{psu}(2,2|4)\) symmetry algebra. To properly characterize the beta deformation, the first important observation is that the antisymmetric representation \(\mathfrak{g} \wedge \mathfrak{g}\) is not irreducible, as it contains the following invariant subspace:  
\begin{align}\label{def:(g^g)0}
&\bullet\ (\mathfrak{g} \wedge \mathfrak{g})_0 := \Big\{ \sum_I x_I \wedge y_I \in \mathfrak{g} \wedge \mathfrak{g} \ \Big| \ \sum_I [x_I, y_I] = 0 \Big\}.
\end{align}
Denoting by $t_a$ the generators of the adjoint representation $\mathfrak{g}$, we can characterize the above invariant subspace by elements of the form $B^{ab}t_a\wedge t_b$, where $B^{ab}$ is a constant and antisymmetric matrix, satisfying the constraint $B^{ab}f_{ab}^c=0$, with $f_{ab}^c$ the structure constant: $[t_a,t_b]=f_{ab}^ct_c$.

The beta deformation states lie in this invariant subspace. Indeed, in the worldsheet description of the beta deformation, the states living in $\mathfrak{g}\wedge\mathfrak{g}$ and not in $(\mathfrak{g}\wedge\mathfrak{g})_0$ correspond to extra degrees of freedom that do not appear in the supergravity spectrum \cite{andrei-rivelles, Kim}. Similarly, in the \(\mathcal{N} = 4\) SYM theory, these states correspond to nonphysical field configurations \cite{segundo}. The space $(\ref{def:(g^g)0})$ contains another invariant subspace:
\begin{align}\label{ftt=0}
&\bullet\ \mathfrak{g} \subset (\mathfrak{g} \wedge \mathfrak{g})_0 \ \text{spanned by} \ \Big\{ f_a^{\ bc} t_b \wedge t_c \Big\}, \hspace{2.5cm}
\end{align}
which corresponds to pure-gauge states. We therefore should factor out the states living in this representation. The beta deformation is thus defined as the states living in the invariant subspace \((\mathfrak{g} \wedge \mathfrak{g})_0\), modulo the pure-gauge states: 
\begin{equation}\label{rep_betadef}
\text{$\beta$-deformation} = \frac{(\mathfrak{g} \wedge \mathfrak{g})_0}{\mathfrak{g}}.
\end{equation}

The most studied example of this deformation in the context of string theory is the pure spinor formalism in \(AdS_5 \times S^5\). In this case, the deformation is realized through a vertex operator that belongs to the representation described in (\ref{rep_betadef}), and is given by the product of two conserved currents \(j\):  
\begin{equation}
V = B^{ab}j_a\wedge j_b.
\end{equation}
The current \(j = j^a t^{R}_a\) lies in the adjoint representation, \(R = \textit{Ad}\), of the symmetry algebra \(\mathfrak{psu}(2,2|4)\). The matrix \(B^{ab}\) is a constant that specifies the irreducible representation, i.e., \(B^{ab} \in (\mathfrak{g} \wedge \mathfrak{g})_0 / \mathfrak{g}\). 

In the upcoming sections, we will extend this description to the case of the B-model topological string theory on the super-twistor space \(\mathbb{CP}^{3|4}\). This topological string belongs to a class of models described in \cite{Witten-Bmodel}. The B-model on \(\mathbb{CP}^{3|4}\) is particularly intriguing due to its connection with \(\mathcal{N}=4\) SYM theory. Specifically, the spectrum of physical states of the topological string can be mapped, via the Penrose transform, to that of the \(\mathcal{N}=4\) theory \cite{twistor}.  

This twistor string theory also contains a BRST symmetry, with its physical states described as elements of the BRST cohomology. These elements will be referred to as vertex operators, and they can be classified according to a quantum number called \textit{ghost number}.

Throughout this paper, we will study the vertex operators with ghost numbers one and two. Operators with ghost number one will be shown to generate the symmetries of the model, while those with ghost number two will describe the beta deformation within the topological model.  

The approach to achieve these results begins by computing the cohomology of the BRST operator \(Q_{\text{BRST}}\). Subsequently, we apply a gauge-fixing procedure known as the \textit{Siegel gauge}, which ensures that the states under consideration are primary fields on the worldsheet. This involves requiring that the operators have no double poles with the stress tensor. Since the topological model contains a composite \(b\)-ghost — a field satisfying the relation \(Qb = T\), where \(T\) is the stress tensor — the gauge-fixing condition can be reformulated as requiring no single poles with the \(b\)-field: \(b_0 U = 0\) \cite{chandia}. Here, $b_0$ is an operator constructed out of the $b$-fields as:
\begin{equation}
b_0(V) = \oint (zdzb_{zz})V.
\end{equation}
The procedure is summarized below in the Figure.
\begin{center}
\includegraphics[width=.9 \linewidth]{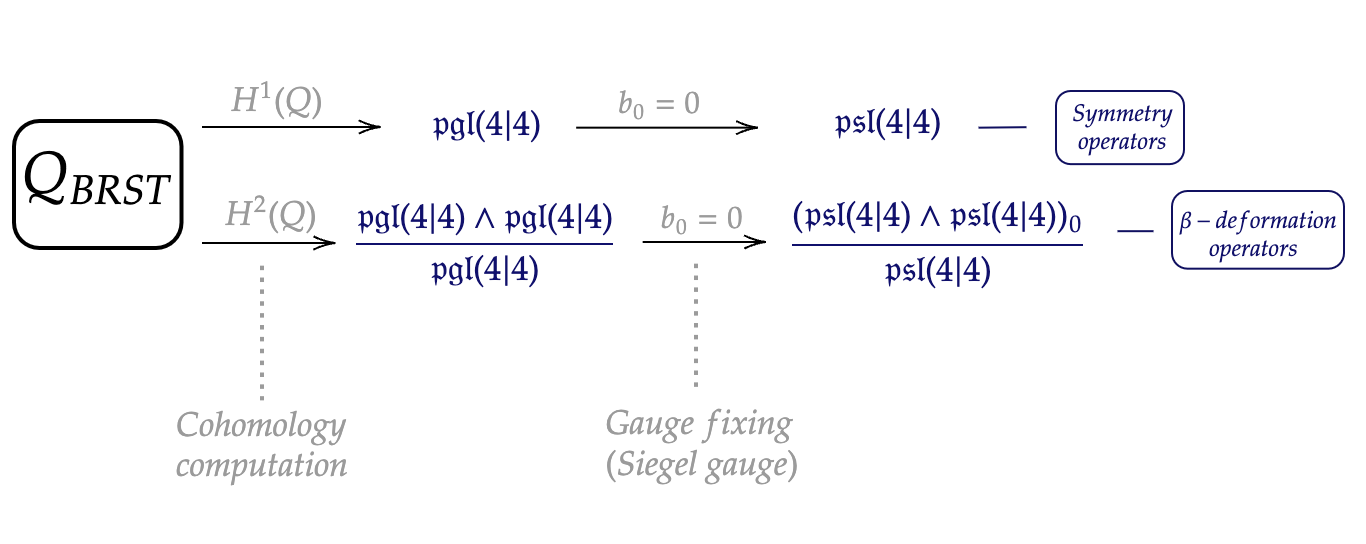}
\end{center}

As a result of this procedure, the beta deformation of the topological model is identified with states in the irreducible representation (\ref{rep_betadef}), which defines the beta deformation. The first step of this procedure is explained in Section 4.2, where it is shown that the ghost number two operators lives in $\frac{(\mathfrak{g}\wedge\mathfrak{g})}{\mathfrak{g}}$. We will see that this quotient space arises naturally from the cohomological computation in projective space, with the denominator justified by Equation (\ref{ftt=0proof}). The second step is discussed in Section 5.2, where, after imposing the Siegel gauge, the states fall within the invariant subspace (\ref{def:(g^g)0}).

Before studying the deformation properly, we will review the topological string in the next section. There, important concepts such as the BRST operator and the $b$-ghost will be introduced.

\section{The B model in projective space}

In this section, we will review the twistor string theory, which is described by the topological B-model on the projective space \(\mathbb{CP}^{3|4}\). The general topological sigma model defined on Calabi-Yau manifolds was first introduced in \cite{WittenTop,Witten-Bmodel}. The case where the target space is the projective supermanifold \(\mathbb{CP}^{3|4}\) has been extensively studied in \cite{twistor,confgra}. 

To describe this model, one introduces fields \(\Phi:\Sigma \longrightarrow \mathbb{CP}^{3|4}\), which are maps from a Riemann surface \(\Sigma\) to the target space. The fields can be expressed by local functions \(\Phi = \Phi^{\mathcal{I}}(z, \bar{z})\) that take values in \(\mathbb{CP}^{3|4}\). The index \(\mathcal{I}\) can be organized as \(\mathcal{I} = (I, A)\), where \(I=1,\ldots,4\) labels the bosonic coordinates \(Z^I\), and \(A=1,\ldots,4\) labels the fermionic coordinates \(\psi^A\). Thus, we have \(\Phi^{\mathcal{I}} = (Z^I, \psi^A)\). The fields are subject to the equivalence relation \(\Phi^{\mathcal{I}} \sim t \Phi^{\mathcal{I}}\) for any non-zero \(t\).

The other fields are a constant Hermitian metric \(g_{\mathcal{I}\bar{\mathcal{I}}}\) and a \(U(1)\) gauge field \(B\). We can now define the covariant derivative acting on \(\Phi\) as 
\[
D\Phi^{\mathcal{I}} = d\Phi^{\mathcal{I}} + iB\Phi^{\mathcal{I}}.
\]
The action is given by:
\begin{equation}\label{model}
S = \frac{1}{2}\int d^2z \ g_{{\cal I}\bar{\cal I}}D_z\Phi^{\cal I}D_{\bar z}\Bar{\Phi}^{\bar{\cal I}}+ \ell(g_{{\cal I}\bar{\cal I}}\Phi^{\cal I}\bar \Phi^{\bar{\cal I}}-r).
\end{equation}
In this action, the lagrangian multiplier $\ell$ introduces the constraint:
\begin{equation}\label{constraintPhi}
g_{{\cal I}\bar{\cal I}}\Phi^{\cal I}\bar \Phi^{\bar{\cal I}}=r.
\end{equation}
This constraint fixes part of the equivalence relation on the fields $\Phi$. The remaining part of the equivalence relation is encoded in the $U(1)$ gauge symmetry:
\begin{equation}
\delta_\alpha\Phi^{\cal I} = e^{i\alpha}\Phi^{\cal I},\ \ \alpha\in[0,2\pi].
\end{equation}

Consider now \(K\) and \(\bar{K}\), the canonical and anti-canonical line bundles of \(\Sigma\). These are the bundles of \(1\)-forms of type \((1,0)\) and \((0,1)\), respectively. In other words, they correspond to the bundles with fibers \(T^{1,0}_p\Sigma\) and \(T^{0,1}_p\Sigma\). The square roots of these bundles are denoted \(K^{1/2}\) and \(\bar{K}^{1/2}\). 

The fermionic fields of the model are denoted by \(\Psi_+^{\cal I}\) and \(\Psi_-^{\cal I}\), which are sections defined in the following spaces:
\begin{align}\label{fermi}
\Psi^{\cal I}_+\in\Gamma(K^{1/2}\otimes\Phi^*( T^{1,0}\mathbb{CP}^{3|4}))\ ,\ \ \
\Psi^{\bar{\cal I}}_+\in\Gamma(K^{1/2}\otimes\Phi^*( T^{0,1}\mathbb{CP}^{3|4})),\\
\Psi^{\cal I}_-\in\Gamma(\bar K^{1/2}\otimes\Phi^*( T^{1,0}\mathbb{CP}^{3|4}))\ ,\ \ \ \Psi^{\bar{\cal I}}_-\in\Gamma(\bar K^{1/2}\otimes\Phi^*( T^{0,1}\mathbb{CP}^{3|4})),\nonumber
\end{align}
In other words, these fields are sections in the projective space, with conformal weights \((1/2,0)\) and \((0,1/2)\) on the worldsheet. Now, we can introduce a fermionic term of the form $\mathcal{L}_D = \Psi_-^{\bar{\cal I}}D_z\Psi_-^{{\cal I}}g_{\bar{\cal I}{\cal I}} + \Psi_+^{\bar{\cal I}}D_{\bar z}\Psi_+^{\cal I}g_{\bar{\cal I}{\cal I}}$ in the Lagrangian. As described in Appendix \ref{ApxProjective}, the variable of the tangent bundle of the projective space satisfies the equivalence relation $\Psi\sim t\Psi$, together with the orthogonality constraint (\ref{constraintbundle}). Therefore, the spinor field $\Psi$ satisfies the following constraint:
\begin{equation}\label{constraintPsi}
g_{{\cal I}\bar{\cal I}}\Phi^{\cal I}\Psi^{\bar{\cal I}}_\pm  = g_{\bar{\cal I}{\cal I}}\bar\Phi^{\bar{\cal I}}\Psi^{\cal I}_\pm = 0.
\end{equation}

We can now define local coordinates, which will eliminate the constraint and the gauge symmetry of the theory defined in global coordinates. With the correct number of degree of freedom, it will be natural to introduce the supersymmetry of the system. Define a local coordinate as:

\begin{equation}
\phi^i := \Phi^i/\Phi^1,\hspace{1cm} \psi^i :=\Psi^i/\Phi^1.
\end{equation}

The fields $\phi^i$ and $\psi^j$ represents the local coordinates of $\mathbb{CP}^{3|4}$ and the tangent bundle $T\mathbb{CP}^{3|4}$, respectively. The Lagrangian now takes the following form:
\begin{align}\label{model}
\begin{split}
S = \int d^2z \Big(g_{i\bar j}(\partial_z\phi^i\partial_{\bar z}\Bar{\phi}^{\bar j} &+ \partial_{\bar z}\phi^i\partial_{z}\Bar{\phi}^{\bar j}) + g_{\bar i j}\psi_-^{\bar i}\nabla_z\psi_-^j+ g_{\bar i j}\psi_+^{\bar i}\nabla_{\bar z}\psi_+^j\\
&+ R_{i\bar ij\bar j}\psi_+^i\psi_+^{\bar i}\psi_-^j\psi_-^{\bar j} \Big),
\end{split}
\end{align}
where $R_{i\bar ij\bar j}$ is the Riemann tensor of the projective space, while $\nabla_z$ and $\nabla_{\bar z}$ are covariant derivatives:
\begin{equation}
\nabla_{z,\bar z}\psi_\pm^I = \partial_{z,\bar z}\psi_\pm^I + \partial_{z,\bar z}\phi^J\Gamma^I_{JK}\psi_\pm^K,
\end{equation}
where $\Gamma^I_{JK}$ is the affine connection on the projective space with $I=i,\bar i$. The supersymmetries of the model are generated by the following infinitesimal transformations: 
\begin{align}\label{transformations}
\delta\phi^i = \alpha_-\psi^i_++\alpha_+\psi^i_-\ &,\ \ \ \delta\phi^{\bar i} = \tilde\alpha_-\psi^{\bar i}_++\tilde\alpha_+\psi^{\bar i}_-, \nonumber\\
\delta\psi^i_+ = -\tilde\alpha_-\partial_z\phi^i\ &,\ \ \ \delta\psi^{\bar i}_+ = -\alpha_-\partial_z\phi^{\bar i},\\
\delta\psi^i_- = -\tilde\alpha_+\partial_{\bar z}\phi^i\ &,\ \ \delta\psi^{\bar i}_- = -\alpha_+\partial_{\bar z}\phi^{\bar i},\nonumber
\end{align}
where $\alpha_-,\tilde\alpha_-\in\Gamma(K^{-1/2})$ are holomorphic sections of $K^{-1/2}$, and $\alpha_+,\tilde\alpha_+\in\Gamma(\bar K^{-1/2})$  are anti-holomorphic sections of $\bar K^{-1/2}$.
From this general sigma model, the topological $B$-model is constructed by twisting the fermionic fields. This twist consists of modifying the conformal weight of the fields, by considering $\psi^{\bar i}_\pm$ as sections of $\Phi^*(T^{0,1}\mathbb{CP}^{3|4})$, and $\psi^{i}_\pm$ as sections of $K\otimes\Phi^*(T^{0,1}\mathbb{CP}^{3|4})$:
\begin{align}
\psi^{\bar i}_+\longmapsto\Gamma\big(\Phi^*(T^{0,1}\mathbb{CP}^{3|4})\big)&,\ \ \ \psi^{\bar i}_-\longmapsto\Gamma\big(\Phi^*(T^{0,1}\mathbb{CP}^{3|4})\big),\\
\psi^i_+\longmapsto\Gamma\big(K\otimes\Phi^*(T^{1,0}\mathbb{CP}^{3|4})\big)&,\ \ \ \psi^{i}_-\longmapsto\Gamma\big(\bar K\otimes\Phi^*(T^{1,0}\mathbb{CP}^{3|4})\big),
\end{align}
In this case, the fields $\psi_\pm^{\bar i}$ have zero conformal weight, while $\psi_+^i$ and $\psi_-^i$ have conformal weight $(1,0)$ and $(0,1)$. Now, it is convenient to combine fields into new ones with zero and $(1,1)$ conformal weights and in the following \cite{twistor,Witten-Bmodel}:
\begin{align}
&\eta^{\bar i} = \psi^{\bar i}_+ + \psi^{\bar i}_-,\\
&\theta_i = g_{i\bar i}\left(\psi^{\bar i}_+ - \psi^{\bar i}_-\right),\\
&\rho^i = \rho^i_zdz + \rho^i_{\bar z}d\bar z,
\end{align}
with $\rho^i_z=\psi_+^i$ and $ \rho^i_{\bar z}=\psi_-^i$. The action in local coordinates turns:
\begin{align}\label{actiontwisted}
\begin{split}
S = \int_\Sigma d^2z\Big(&g_{i\bar j}(\partial_z\phi^i\partial_{\bar z}\Bar{\phi}^{\bar j} + \partial_{\bar z}\phi^i\partial_{z}\Bar{\phi}^{\bar j}) +\bar\eta^{\bar i}(\nabla_z\rho_{\bar z}^i+\nabla_{\bar z}\rho_z^i)g_{i\bar i}\\
&\ \ \ + \theta_i(\nabla_{\bar z}\rho_{z}^i-\nabla_{z}\rho_{\bar z}^i)+R_{i\bar ij\bar j}\rho^i\rho^j\eta^{\bar i}\theta^{\bar j}\Big).
\end{split}
\end{align}
The topological transformation laws are found from (\ref{transformations}) by setting $\alpha_\pm=0$ and $\tilde\alpha_+=\tilde\alpha_-=\alpha= constant$:
\begin{align}
& Q \phi^i = 0\ ,\ \ Q\bar \phi^{\bar i} = \alpha\eta^{\bar i},\\
&Q\eta^{\bar i} = Q\theta_{i}=0,\\
&Q\rho_z^i=-\alpha\partial_z\phi^i\ ,\ \ \ Q\rho_{\bar z}^i=-\alpha\partial_{\bar z}\phi^i.
\end{align}
The B-model is a topological theory, in that being independent of the worldsheet metric and on the Kähler metric $g_{i\bar i}$. We can readily see this by rewriting the action as \cite{Hofman}:
\begin{align}\label{actiontwisted}
\begin{split}
S_B = \int_\Sigma d^2z\ & \rho^id\theta_i + Q\Big(\rho_z^i\partial_{\bar z}\bar \phi_i+\rho_{\bar z}^i\partial_z\bar \phi_i - \frac{1}{2}\Gamma^i_{jk}\rho^j\rho^k\theta_i\Big)\\
 = \int_\Sigma d^2z\ \ &(\partial_z\phi^i\partial_{\bar z}\Bar{\phi}_i + \partial_{\bar z}\phi^i\partial_{z}\Bar{\phi}_i) + (\rho^i_{\bar z}\partial_{z}\theta_i - \rho^i_{z}\partial_{\bar z}\theta_i)\\
&\ \ \ \ \ +(\rho^i_{\bar z}\nabla_{z}\eta_i + \rho^i_{z}\nabla_{\bar z}\eta_i) - Q\Big(\frac{1}{2}\Gamma^i_{jk}\rho^j\rho^k\theta_i\Big),
\end{split}
\end{align}
from which we can read the following OPE's:
\begin{align}
\begin{split}
&\phi^i(z)\bar \phi_j(0) \sim \frac{\delta^i_j}{2}\ln|z|^2,\\
&\theta_j(z)\rho_z^i(0)\sim\frac{\delta^i_j}{z}\ ,\ \ \ \theta_j(z)\rho_{\bar z}^i(0)\sim-\frac{\delta^i_j}{\bar z}.
\end{split}
\end{align}

The energy-momentum tensor is defined as the variation of the Lagrangian with respect to the metric, and it reads:
\begin{equation}
T_{zz} = \partial_z\phi^i\partial_{z}\bar\phi_i + \rho^i\nabla_z\eta_i\ ,\ \ \ T_{\bar z\bar z} = \partial_{\bar z} \phi_z^i\partial_{\bar z}\bar \phi_i + \rho_{\bar z}^i\nabla_{\bar z}\eta_i.
\end{equation}
Therefore, from the analysis of the action we did in the previous paragraph, we conclude that the energy momentum tensor is BRST-exact. Then, there should be a field $b$, such that $Qb=T$. This is the definition of the $b$-ghost, which is given by \cite{Witten-CS}:
\begin{equation}\label{b-ghost}
b_{zz} = \rho^i_z\partial_z\bar \phi_i,\ \ \ \ \ b_{\bar z\bar z} = \rho^i_{\bar z}\partial_{\bar z}\bar \phi_i.
\end{equation}

\noindent One can eventually return to the global coordinates, by defining the following relation:
\begin{align}
&H^{\bar{\cal I}} = \Psi^{\bar{\cal I}}_+ + \Psi^{\bar{\cal I}}_-,\\
&\Theta_{\cal I} = g_{{\cal I}\bar{\cal I}}\left(\Psi^{\bar{\cal I}}_+ - \Psi^{\bar{\cal I}}_-\right),\\
&P^{\cal I} = P^{\cal I}_zdz + P^{\cal I}_{\bar z}d\bar z,
\end{align}
with $P^{\cal I}_z=\Psi_+^{\cal I}$ and $ P^{\cal I}_{\bar z}=\Psi_-^{\cal I}$. In such a way that:
\begin{equation}
\theta_i = \Phi^1\Theta_i\ ,\ \ \ \ \eta^i = H^i/\Phi^1\ ,\ \ \ \ \rho^{i} = P^{i}/\Phi^1.
\end{equation}
These global coordinates parametrizes the fiber of the tangent bundle (see Appendix \ref{ApxProjective}), and therefore satisfies the following constraints:
\begin{equation}
\Phi^{\cal I}\Theta_{\cal I} = 0\ ,\ \ \ \ g_{{\cal I}\bar{\cal I}}\Phi^{\cal I}H^{\bar {\cal I}} = 0\ ,\ \ \ \ g_{\bar{\cal I}{\cal I}}\bar\Phi^{\bar I}P^{{\cal I}} = 0.
\end{equation}

An interesting observation is that the $b$-ghost is well defined in the global coordinates, as:
\begin{equation}
b = \rho^{i}d\bar\phi_{i} = P^{i}d\bar\Phi_{i} - P^{i}\bar\Phi_{i}\frac{d\bar\Phi_0}{\bar\Phi_0}  = P^{i}d\bar\Phi_i + P^0\bar\Phi_0\frac{d\bar\Phi_0}{\bar\Phi_0} = P^{\cal I}d\bar\Phi_{\cal I}.
\end{equation}

\section{Vertex operators}

In this section, we will describe the vertex operators of the topological B-model in $\mathbb{CP}^{3|4}$. These operators are elements of the BRST cohomology, with the BRST operator defined in the last section:
\begin{equation}
Q_{B} = \eta^{\bar i}\frac{\partial}{\partial\bar\phi^{\bar i}} + d\phi^i\frac{\partial}{\partial\rho^i}.
\end{equation}

The cohomology classes of this operator are represented by operators that are functions of the fields $\phi^i,\bar\phi^{\bar i},\theta_i,\eta^{\bar i}$, with the form:

\begin{equation}\label{V}
\mathcal{O}_V = \eta^{\bar i_1}\cdots \eta^{\bar i_p}V_{\bar i_1\cdots\bar i_p}^{\ \ \ \ \ j_1\cdots j_q}\theta_{j_1}\cdots\theta_{j_q},
\end{equation}
such that $V$ is holomorphic, $\bar\partial V = 0$, and $V\ne\bar\partial \Lambda$. We can interpret $\eta^{\bar i}$ as the $(0,1)$-forms $d\bar\phi^{\bar i}$, and $\theta_i$ as the $(1,0)$-tangent vectors $\frac{\partial}{\partial \phi^i}$. In this case, the vertex operators (\ref{V}) will be interpreted as holomorphic $(0,p)$-forms on the projective space, with values in $\wedge^qT{\mathbb{CP}^{3|4}}$, the antisymmetric product of $q$ holomorphic tangent bundle $T\mathbb{CP}^{3|4}$. Explicitly:
 \begin{equation}
V= d\bar\phi^{\bar i_1}\cdots d\bar\phi^{\bar i_p}V_{\bar i_1\cdots\bar i_p}^{\ \ \ \ \ j_1\cdots j_q}\frac{\partial}{\partial\phi^{j_1}}\cdots\frac{\partial}{\partial\phi^{j_q}}.
\end{equation}

With this interpretation, the BRST operator $Q$ is identified with the $\bar\partial$ operator acting on these forms. Therefore, $V$ is an element of the $\bar\partial$ cohomology group $H^p\left(\mathbb{CP}^{3|4},\wedge^qT{\mathbb{CP}^{3|4}}\right)$. Then, by identifying the operators $\mathcal{O}_V$ and $V$, the physical states consists of the direct sum:
\begin{equation}\label{physicalstates}
\text{Physical states}\ \cong\ \bigoplus_{p,q}H^p\left(\mathbb{CP}^{3|4},\wedge^qT{\mathbb{CP}^{3|4}}\right).
\end{equation}

These cohomology groups are finite-dimensional since the projective space is compact. For a suitable open subset of the projective space, the cohomology groups of $\bar\partial$ have infinite dimension and can be identified with plane wave solutions \cite{twistor}. A remarkable example is exemplified in \cite{confgra}, where a subsector of these vertex operators are related to the conformal supergravity fields.

Our goal here is to study the physical states defined on the complete projective space, without restricting to any specific subset. In this case, the physical states will live in a finite-dimensional representation of the symmetry algebra of the string model. To properly study these states, we introduce a quantum number called the \textit{ghost number}. The ghost number is defined as follows: it is $1$ for $\eta$ and $\theta$, $-1$ for $\rho$, and $0$ for $\phi$. Accordingly, $Q$ is a ghost number $1$ operator. The states defined in equation (\ref{physicalstates}) have a ghost number given by the sum $p + q$. Therefore, we can compute the physical states for each ghost number. In particular, we are interested in the states of ghost number one and two:

\begin{equation}\label{Hpq}
\bigoplus_{p+q=i}H^p\left(\mathbb{CP}^{3|4},\wedge^qT{\mathbb{CP}^{3|4}}\right),\ \ \ i=1,2.
\end{equation}
It turns out that the only non-vanishing cohomologies above are given when $p=0$ \cite{Noja}. Therefore, if we are interested in the states with ghost number $i=1$ and $2$, we have to compute the following cohomology group:
\begin{equation}\label{H0}
H^0\left(\mathbb{CP}^{3|4},\wedge^iT{\mathbb{CP}^{3|4}}\right).
\end{equation}

\subsection{Ghost number 1 operators}\label{1ghost}

The ghost number $1$ operators will be obtained by computing the set $H^0\left(\mathbb{CP}^{3|4},T\mathbb{CP}^{3|4}\right)$, which is the set of global holomorphic vector fields in the projective space. In terms of local coordinates $\phi^i$, such vector fields are of the form:
\begin{equation}
V(\phi)=V^i(\phi)\frac{\partial}{\partial\phi^i}.
\end{equation}
Being globally defined means that the vector field can be evaluated at any point on projective space and behaves consistently under coordinate transformations. Consequently, it can be expressed in terms of local coordinates at any given point. However, a more straightaway way to define a global vector field is by writing it in terms of the global coordinates $\Phi^{\cal I}$:
\begin{equation}
V(\Phi) = V^{\cal I}(\Phi)\frac{\partial}{\partial\Phi^{\cal I}},
\end{equation}
with the condition that it descends to the projective space, which means that it is invariant under rescaling of the global coordinates: $V(t\Phi) = V(\Phi)$. As explained in Appendix \ref{ApxProjective}, under rescaling, the basis of the vector fields in the projective space transforms as ${\partial}/{\partial\Phi^{\cal I}} \mapsto t^{-1} {\partial}/{\partial\Phi^{\cal I}}$. Therefore, the coefficients of the vector field should transforms as $V^{\cal I}(\Phi)\mapsto V^{\cal I}(t\Phi) = tV^{\cal I}(\Phi)$. The expressions that satisfies this are the linear functions of $\Phi$: 
$V^{\cal I}(\Phi) = v^{\cal I}_{\cal J}\Phi^{\cal J}$. As also explained in the Appendix \ref{ApxProjective}, the vector fields written in global coordinates are submitted to the condition $\Phi^{\cal I}\frac{\partial}{\partial\Phi^{\cal I}}=0$. In conclusion, the set of global sections in the projective space is given by the expressions of the form:
\begin{equation}\label{Vg1}
V(\Phi) = v^{\cal I}_{\cal J}\Phi^{\cal J}\frac{\partial}{\partial\Phi^{\cal I}},
\end{equation}
with the condition $\Phi^{\cal I}\frac{\partial}{\partial\Phi^{\cal I}}=0$. In conclusion, the global vector fields are given by the linear operators on the projective space $\mathbb{CP}^{3|4}$. In other words, it is given by the set of super-matrices of dimension $(4|4)\times(4|4)$, factored out by the identity operator $\textbf{1}=\Phi^{\cal I}\frac{\partial}{\partial\Phi^{\cal I}}$, which forms the lie projective superalgebra $\mathfrak{pgl}(4|4)=\mathfrak{gl}(4|4)/\mathfrak{u}(1)$.

We can now write the vertex operator of ghost number $1$, based on the identification of $\frac{\partial}{\partial\Phi^{\cal I}}$ with $\Theta_{\cal I}$:
\begin{equation}\label{vertexone}
\mathcal{O}_V = v^{\cal I}_{\cal J}\Phi^{\cal J}\Theta_{\cal I},
\end{equation}
Together with the condition $\Phi^{\cal I}\Theta_{\cal I}=0$. Therefore, we conclude that the ghost number 1 vertex operators are representations of the $\mathfrak{pgl}(4|4)$ algebra:
\begin{equation}
H^1(Q_{B}) \cong \mathfrak{pgl}(4|4).
\end{equation}

\subsection{Ghost number $2$ operators}

For the ghost number 2 operators, we repeat the same logic. We now need to compute $H^0\left(\mathbb{CP}^{3|4},\wedge^2T\mathbb{CP}^{3|4}\right)$, which is the set of global holomorphic sections of antisymmetric bi-vector fields. In global coordinates, it takes the form:
\begin{equation}
W(\Phi) =  W^{\mathcal{I}\mathcal{J}}(\Phi)\frac{\partial}{\partial\Phi^{\cal I}}\wedge\frac{\partial}{\partial\Phi^{\cal J}},
\end{equation}
with the condition $\Phi^{\cal I}\frac{\partial}{\partial\Phi^{\cal I}}=0$. Again, we must impose the condition that the section descend to the projective space, which means that $W(t\Phi)=W(\Phi)$. Under this rescaling, the basis of the antisymmetric bi-vector fields transforms as $(\partial/\partial\Phi^{\cal I}\wedge\partial/\partial\Phi^{\cal J})\mapsto t^{-2} (\partial/\partial\Phi^{\cal I}\wedge\partial/\partial\Phi^{\cal J})$. Therefore, the coefficient of the section must be a polynomial of degree two. Then, the global sections we are interested in are of the following form:
\begin{equation}
W(\Phi) =  B^{\cal I\cal J}_{\cal K\cal L}\Phi^{\cal K}\Phi^{\cal L}\frac{\partial}{\partial\Phi^{\cal I}}\wedge\frac{\partial}{\partial\Phi^{\cal J}}.
\end{equation}

As in the previous case, we can also think of this operator as antisymmetric tensors in the projective space, that is, the antisymmetric product of the projective superalgebra:
\begin{equation}
W = B^{\cal I\cal J}_{\cal K\cal L}\ \Phi^{\cal K}\frac{\partial}{\partial\Phi^{\cal I}}\wedge\Phi^{\cal L}\frac{\partial}{\partial\Phi^{\cal J}}\in\mathfrak{pgl}(4|4)\wedge\mathfrak{pgl}(4|4).
\end{equation}

This antisymmetric product of the two projective algebras still does not characterize the cohomology group. Indeed, we still need to impose the condition that $\Phi^{\cal I}\frac{\partial}{\partial\Phi^{\cal I}}=0$ to the tensor fields. The vectors of the form $\Phi^{\cal I}\frac{\partial}{\partial\Phi^{\cal I}}\wedge\Phi^{\cal L}\frac{\partial}{\partial\Phi^{\cal J}}$ are already factored out. However, we still need to factorize it by the vectors of the form:
\begin{equation}
\Phi^{\cal K}\frac{\partial}{\partial\Phi^{\cal M}}\wedge\Phi^{\cal N}\frac{\partial}{\partial\Phi^{\cal K}}.
\end{equation}

We can characterize this set as an irreducible representation of the projective algebra, defined in (\ref{ftt=0}). In order to see this, denote by $t^{\cal I}_{\cal J}=\Phi^{\cal I}\frac{\partial}{\partial\Phi^{\cal J}}$ the basis of the algebra. Then, using the structure constant $f_{^{\mathcal{I}\mathcal{K}} _{\mathcal{J}\mathcal{L}}}^{\ \ \ ^\mathcal{M}_\mathcal{N}}$ in (\ref{f}), we see that:
\begin{align}\label{ftt=0proof}
\begin{split}
f_{^{\mathcal{I}\mathcal{K}} _{\mathcal{J}\mathcal{L}}}^{\ \ \ ^\mathcal{M}_\mathcal{N}}t^\mathcal{I}_\mathcal{J}\wedge t^\mathcal{K}_\mathcal{L} & = \Big(\delta^\mathcal{K}_\mathcal{J}\delta^\mathcal{I}_\mathcal{M}\delta^\mathcal{N}_\mathcal{L}- (-1)^{\left|^\mathcal{I}_\mathcal{J}\right|\cdot\left|^\mathcal{K}_\mathcal{L}\right|}\delta_\mathcal{L}^\mathcal{I}\delta^\mathcal{K}_\mathcal{M}\delta^\mathcal{N}_\mathcal{J}\Big)\Phi^{\cal J}\frac{\partial}{\partial\Phi^{\cal I}}\wedge\Phi^{\cal L}\frac{\partial}{\partial\Phi^{\cal K}}\\
& = \Phi^{\cal K}\frac{\partial}{\partial\Phi^{\cal M}}\wedge\Phi^{\cal N}\frac{\partial}{\partial\Phi^{\cal K}} - (-1)^{\left|^\mathcal{I}_\mathcal{J}\right|\cdot\left|^\mathcal{K}_\mathcal{L}\right|}\Phi^{\cal N}\frac{\partial}{\partial\Phi^{\cal L}}\wedge\Phi^{\cal L}\frac{\partial}{\partial\Phi^{\cal M}}\\
& = 2\Phi^{\cal K}\frac{\partial}{\partial\Phi^{\cal M}}\wedge\Phi^{\cal N}\frac{\partial}{\partial\Phi^{\cal K}}.
\end{split}
\end{align}

In conclusion, the cohomology group we are studying is given by the antisymmetric product of two projective algebras, factored out by the algebra generated by elements of the form $f_{ab}^{\ \ c}t^a\wedge t^b$. This can be summarized in the following factor space: $\frac{\mathfrak{pgl}(4|4)\wedge\mathfrak{pgl}(4|4)}{\mathfrak{pgl}(4|4)}$.

We can now write the vertex operator of ghost number $2$, again by identifying the vector fields $\frac{\partial}{\partial\Phi^{\cal I}}$ with the fields $\Theta_{\cal I}$:
\begin{equation}\label{ghost2}
\mathcal{O}_W =  (B^{\cal I\cal J}_{\cal K\cal L}\Phi^{\cal K}\Phi^{\cal L})\Theta_{\cal I}\Theta_{\cal J},
\end{equation}
where $B^{\cal I\cal J}_{\cal K\cal L}\in\frac{\mathfrak{pgl}(4|4)\wedge\mathfrak{pgl}(4|4)}{\mathfrak{pgl}(4|4)}$ means that it is an antisymmetric tensor: $ B^{\mathcal{I}\mathcal{K}}_{\mathcal{J}\mathcal{L}} = - B^{\mathcal{K}\mathcal{I}}_{\mathcal{L}\mathcal{J}}$, living in the antisymmetric product $\mathfrak{pgl}(4|4)\wedge\mathfrak{pgl}(4|4)$, and satisfying the equivalence relation:
\begin{equation}\label{equiv.rel}
B^{\cal I\cal J}_{\cal K\cal L}\sim B^{\cal I\cal J}_{\cal K\cal L} +f_{^{\cal I\cal J}_{\cal K\cal L}}^{\ \ \ ^{\cal M}_{\cal N}}A^{\cal M}_{\cal N},
\end{equation}
for any matrix $A$. Therefore, we conclude that the ghost number 2 vertex operators are in the following representations of the projective algebra:
\begin{equation}
H^2(Q_{B}) \cong \frac{\mathfrak{pgl}(4|4)\wedge\mathfrak{pgl}(4|4)}{\mathfrak{pgl}(4|4)}.
\end{equation}

\subsection{Ghost number $n$ operators}

For the general case of ghost number $n$, the operators which are BRST closed and not exact and also invariant under scaling of coordinates are of the form:
\begin{equation}
t^{(\mathcal{I}_1\cdots\mathcal{I}_n)}_{[\mathcal{J}_1\cdots\mathcal{J}_n]} = (\Phi^{\mathcal{I}_1}\Theta_{\mathcal{J}_1})(\Phi^{\mathcal{I}_2}\Theta_{\mathcal{J}_2})\cdots(\Phi^{\mathcal{I}_n}\Theta_{\mathcal{J}_n}),
\end{equation}
together with the condition $\Phi^\mathcal{I}\Theta_\mathcal{I}=0$.

\section{Local coordinates and gauge fixing}

In this section, we will study the vertex operators described in the previous section within a local gauge known as the Siegel gauge \cite{chandia}. Our main result will be the characterization of the ghost number 2 operators as the beta deformation supermultiplets. Additionally, we will interpret the ghost number 1 operators as the symmetry generators of the theory.

The operators described in the previous section contain a gauge ambiguity. This ambiguity arises because some of the operators are not primary and can, therefore, be derived from the primary ones. A primary operator can be characterized as having no double poles with the energy momentum tensor $T$ of the worldsheet theory. Since the energy-momentum tensor is the BRST transformation of the $b$-ghost, $Qb=T$, the condition for primary can be defined as having no single poles with the $b$-ghost. In this section, we will eliminate the operators having non-vanishing single poles with the $b$-ghost. As a result, the remaining operators must satisfy the constraint of a zero internal commutation relation.

To compute these relations, we will work in local coordinates:
\begin{equation}
\Phi^i \longmapsto \phi^i \equiv \frac{\Phi^i}{\Phi^1},\ \ \ \ \ \ \Theta_i \longmapsto \theta_i \equiv \Phi^1\Theta_i,
\end{equation}
such that:
\begin{align}
\begin{split}
& \Phi^1\Theta_1 = - \Phi^i\Theta_i = -\phi^i\theta_i, \hspace{2cm} \Phi^1\Theta_i  = \theta_i,\\
& \Phi^i\Theta_1 = \phi^i\Phi^1\Theta_1 = -\phi^i\phi^k\theta_k, \hspace{1.45cm} \Phi^i\Theta_j = \phi^i\theta_j.
\end{split}
\end{align}
Then, in local coordinates the ghost number one operator $t^\mathcal{I}_\mathcal{J} = \Phi^\mathcal{I}\Theta_\mathcal{J}$ is broken into three components, $t^\mathcal{I}_\mathcal{J} = (t_a, t^{\hat a}, t^i_j)$, given by:
\begin{align}\label{basisghostn1}
\begin{split}
\hspace{3cm}&t_a=\theta_a \\
&t^{\hat a} = \phi^{\hat a}\phi^j\theta_j\\
&t^i_j = \phi^i\theta_j, \hspace{1cm} i,j,a,\hat a=2,\cdots,8.
\end{split}
\end{align}

We can also compute the infinitesimal generators of the projective algebra, as explained in Appendix \ref{ApxProjective2}. In this case, the structure constants will be:
\begin{align}
\begin{split}
&f_{ ^{ik} _{jl}}^{\ \ ^m_n} = \delta^k_j\delta_m^i\delta^n_l- (-1)^{|^i_j|\cdot |^k_l|}\delta^i_l\delta_m^k\delta^n_j,\\
&f_{a\hat a}^{\ \ ^m_n} =  \delta^{\hat a}_m\delta^n_a + (-1)^{|a|\cdot|\hat a|} \delta^{\hat a}_a\delta^n_m,\hspace{3cm} f_{\hat aa}^{\ \ ^m_n}=-(-1)^{|a|\cdot|\hat a|}f_{a\hat a}^{\ \ ^m_n},\\
&f_{a\ ^i_j}^{\ \ \ b} = \delta^i_a\delta^b_j,\hspace{6.0cm} f_{^i_j\ a}^{\ \ \ b}= -(-1)^{|a|\cdot|^i_j|}f_{a\ ^i_j}^{\ \ \ b},\\
&f_{\hat a\ ^i_j}^{\ \ \ \hat b}= -\delta^{\hat a}_j\delta^i_{\hat b},\hspace{5.7cm} f_{^i_j\ \hat a}^{\ \ \ \hat b}= -(-1)^{|\hat a|\cdot|^i_j|}f_{\hat a\ ^i_j}^{\ \ \ \hat b}.
\end{split}
\end{align}

In order to compute the operators in the Siegel gauge, we will use the OPE that were defined in section 2:
\begin{align}
\begin{split}
&\phi^i(z)\bar \phi_j(0) \sim \frac{\delta^i_j}{2}\ln|z|^2,\\
&\theta_j(z)\rho_z^i(0)\sim\frac{\delta^i_j}{z}\ ,\ \ \ \theta_j(z)\rho_{\bar z}^i(0)\sim-\frac{\delta^i_j}{\bar z}.
\end{split}
\end{align}

We are now able to define the action of the $b$-ghost (\ref{b-ghost}) by the usage of the local coordinates and the corresponding OPEs above. In the next two subsections, we will express the operators in local coordinates and derive insightful conclusions from the gauge-fixing procedure.

\subsection{Ghost number 1 operators generates symmetry}\label{ghostnumber1}

In local coordinates, the ghost number 1 vertex operator is:
\begin{align}
\begin{split}
V_1 = A^j_i \phi^i\theta_j + A^a\theta_a + A_{\hat a}\phi^{\hat a}\phi^i\theta_i.
\end{split}
\end{align}
Where $A$ are constant coefficients. One can now choose the Siegel gauge, i.e. the gauge where $b_0V_1=0$, as explained earlier. The expression for the $b$-ghost is given in (\ref{b-ghost}). The gauge fixed equations will result in the traceless condition for the operator:
\begin{align}
\begin{split}
0 = b_0(V_1) & = \oint (zdzb_{zz})V_1(0) \\
& = \oint (zdz\wick{\rho^m_z\partial_z\bar\phi_m)(A^j_i \phi^i\theta_j + A^a\theta_a + A_{\hat a}\phi^{\hat a}\phi^i\theta_i)(0)}\\
& = \oint (zdz\wick{\c1\rho^m_z\partial_z\c2{\bar{\phi}}_m)(A^j_i \c2\phi^i\c1\theta_j})(0) + \oint (zdz\wick{\c1\rho^m_z\partial_z{\bar{\phi}}_m)(A^a\c1\theta_a})(0)\\
&\hspace{1cm} + \oint (zdz\wick{\c1\rho^m_z\partial_z\c2{\bar{\phi}}_m)(A_{\hat a}\phi^{\hat a}\c2\phi^i\c1\theta_i)}(0) + \oint (zdz\wick{\c1\rho^m_z\partial_z\c2{\bar{\phi}}_m)(A_{\hat a}\c2\phi^{\hat a}\phi^i\c1\theta_i)}(0)\\
& = A^i_i + 0 - (A\cdot\phi) + (A\cdot\phi) = A^i_i = \Str(A).
\end{split}
\end{align}

In conclusion, after gauge fixing it becomes manifest that the ghost number 1 operators are representations of the $\mathfrak{psl}(4|4)$ algebra. This is an important fact since $PSL(4|4)$ is the real form of the symmetry group $PSU(2,2|4)$ of $\mathcal{N}=4$ SYM. As pointed out in \cite{twistor}, the $PSL(4|4)$ is the group of a global symmetry of topological B-model in the projective space. We will now show that the ghost number 1 operators can be understood as the symmetry generators of this model. For this purpose, define the action of the $V_1$ operator on arbitrary functions $F(\phi,\bar\phi)$ defined in the space of fields as \cite{Lian}:
\begin{equation}
\Vec{V}_1F = b_0 (V_1\cdot F) = \oint zdz(\rho^k_z\partial_z\bar\phi_k)(z)\oint \frac{dw}{w}V_1(w)F(0).
\end{equation}
The result of this computation gives:

\begin{equation}
b_0 (V_1\cdot F) = \left(A^i_j\phi^j\frac{\partial}{\partial\phi^j} + A^a\frac{\partial}{\partial\phi^a} +  A_{\hat a}\phi^{\hat a}\phi^i\frac{\partial}{\partial\phi^i}\right)F,\ \ \ \ A \in \mathfrak{psl}(4|4).
\end{equation}
Which is the infinitesimal action of the $PSL(4|4)$ symmetry group on arbitrary functions. In terms of the super-conformal algebra, the first term above contains the rotations, the second term contains translations and the last term the special conformal transformations.

\subsection{Ghost number 2 operators and beta deformation}

Now, we will see how the ghost number two operators can be understood as the beta deformation states. In the next section, we will use these operators to deform the action and the BRST operator of the twistor string. The ghost number 2 operators (\ref{ghost2}) will take the following form in local coordinates:

\begin{align}\label{V2}
\begin{split}
V_2 = &\ \ B^{ij}_{kl}\phi^k\phi^l\theta_i\theta_j + B^{ia}_{j}\phi^j\theta_i\theta_a + B^{i}_{j\hat a}\phi^j\phi^{\hat a}\phi^k\theta_i\theta_{k}\\
& + B_{\hat a\hat b}\phi^{\hat a}\phi^i\phi^{\hat b}\phi^j\theta_i\theta_j + B^{\ a}_{\hat a}\phi^i\phi^{\hat a}\theta_i\theta_a + B^{ab}\theta_a\theta_b.
\end{split}
\end{align}

Where $B$ are constant antisymmetric matrices in the adjoint representation of the symmetry algebra. As we saw in the previous section, the ghost number 2 vertex that comes from the BRST cohomology is in the following finite dimensional representation of the symmetry algebra:
\begin{equation}\label{pgl^pgl/pgl}
[B]\in\frac{\mathfrak{pgl}\wedge\mathfrak{pgl}}{\mathfrak{pgl}}.
\end{equation}
Here, the brackets in $[B]$ means that we are considering the equivalence class (\ref{equiv.rel}), where the matrix $B$ is a representative of this class. When we go to local coordinates, this equivalence classes are projected into a representative element, which are written in (\ref{V2}).

We can decompose the representation (\ref{pgl^pgl/pgl}) and write it in terms of the symmetry algebra $\mathfrak{psl}$. We first notice that $\mathfrak{pgl} = \mathfrak{psl}\oplus \textbf{J}$, where $\textbf{J}$ is the super-trace part of the algebra: $\textbf{J}=\text{diag}(1,1,1,1,-1,-1,-1,-1)$. Using this, we can do the following decomposition:
\begin{equation}\label{oplus}
B\in\mathfrak{pgl}\wedge\mathfrak{pgl} = (\mathfrak{psl}\wedge\mathfrak{psl})\oplus(\textbf{J}\otimes\mathfrak{psl}),
\end{equation}
which can also be written as:
\begin{equation}
(\textbf{63}\wedge\textbf{63}) = (\textbf{62}\wedge\textbf{62})\oplus\textbf{62}.
\end{equation}

Next, we will find the states of the beta-deformation based on the observations of this section. For this purpose, we will compute the constraint that a gauge fixing imposes to the ghost number two operators.

\subsubsection{Beta deformation vertice operators}

We can now impose the gauge fixing condition on the ghost number two vertex operator by requiring that the operator vanishes under the action of the zero mode of the $b$-ghost. In the following, we will carry out this computation carefully. There are two types of ghost number two operators, as described earlier by equation (\ref{oplus}). The first type consists of the \textit{multiplets} living in the antisymmetric product of two adjoint representations of the symmetry algebra, while the second type consists of the \textit{singlets} living in the adjoint representation of the Lie algebra. The beta deformation states are identified with those in the first type. To see this, we begin by writing a basis for the operators in the $\mathfrak{pgl}\wedge\mathfrak{pgl}$ representation:
\begin{align}\label{localghost2}
\begin{split}
&U_{ab} = \theta_a\wedge\theta_b, \\
&U_{a}^{\ \hat a} = \theta_a\wedge(\phi^{\hat a}\phi\cdot\theta),\\
&U^{\hat a\hat b} = (\phi^{\hat a}\phi\cdot\theta)\wedge(\phi^{\hat b}\phi\cdot\theta),\\
&U^i_{j a} = \phi^i\theta_j\wedge\theta_a ,\\
&U^{i \hat a}_{j}=\phi^i\theta_j\wedge(\phi^{\hat a}\phi\cdot\theta) ,\\
&U^{ik}_{jl} = \phi^i\theta_j\wedge\phi^k\theta_l.
\end{split}
\end{align}
Using this language, the most general ghost number two operator is written as:
\begin{align}
\begin{split}
V_2 = &\ \ B^{ij}_{kl}U^{kl}_{ij} + B^{ia}_{j}U^j_{i a} + B^{i}_{j\hat a}U^{j \hat a}_{i } + B_{\hat a\hat b}U^{\hat a\hat b} + B^a_{\ \hat a}U_a^{\ \hat a} + B^{ab}U_{ab}.
\end{split}
\end{align}

Now, we want to restrict ourselves to the $\mathfrak{psl}\wedge\mathfrak{psl}$ multiplet. For that purpose, we must impose the traceless condition on $B$:
\begin{align}\label{tracelessB}
B^{ij}_{il} = B^{ij}_{kj} = B^{ia}_{i} = B^{i}_{i\hat a} = 0.
\end{align}
where the repeated indices here means super-trace: $B^i_i = \Str(B)$.

The gauge fixing condition will be imposed on each operator in (\ref{localghost2}) by acting with the zero mode of the $b$-ghost, whose expression is given by (\ref{b-ghost}):
\begin{align}\label{bU=0}
\begin{split}
b_0U^{ik}_{jl} = b_0U^i_{j a} = b_0U^{i \hat a}_{j } = b_0U^{\hat a\hat b} = b_0U_a^{\ \hat a} = b_0U_{ab} = 0.
\end{split}
\end{align}

We now turn to the computation of this gauge fixing condition. The first important observation is that the $b$-ghost acts on these operators in two different ways. It can act on only one of the 
$\mathfrak{psl}$-components; or it can act by mixing the two $\mathfrak{psl}$-components. In the first case, it extracts the super-trace of the operator. For instance, this type of contractions on $B^{jl}_{ik}U^{ik}_{jl}$ will give us:
\begin{equation}
B^{jl}_{ik}\oint zdz\wick{(\c1\rho_z^m\partial_z\c2{\bar{\phi}}_m
)(z)(\c2\phi^i\c1\theta_j\wedge\phi^k\theta_l)(0)} = B^{il}_{ik} \phi^k\theta_l = 0,
\end{equation}
and the same thing happens with the other operators. In conclusion, this type of contraction automatically gives zero and therefore will not contribute to new constraints. The new constraints of the gauge fixing comes when the $b$-ghost acts by mixing the components of the operator. We will see now that this imposes the \textit{vanishing internal commutator} condition on $B$:
\begin{equation}\label{vanishinginternalcommutator}
B^{ji}_{kl}\ f_{^{ji} _{kl}}^{\ \ \ ^m_n}  = B_a^{\ \hat a}f_{a\hat a}^{\ \ ^m_n} = B^{i\ a}_{j}f_{^i_j\ a}^{\ \ \ b} = B_j^{i\ \hat a}f_{^i_j\ \hat a}^{\ \ \ \hat b} = 0.
\end{equation}

Let us see how to obtain these constraints, by imposing the condition (\ref{bU=0}):

\begin{itemize}
    \item  $b_0\Big(B^{ji}_{kl}U^{kl}_{ji}\Big)=0$:
\end{itemize}
\begin{align}
\begin{split}
&\ b_0(\phi^i\theta_k\wedge \phi^j\theta_l) = \oint (zdz\rho_z^i\partial_z\bar \phi_i)(z)(\phi^i\theta_k\wedge \phi^j\theta_l)(0)\\
&= \oint zdz\wick{(\c1\rho_z^m\partial_z\c2{\bar{\phi}}_m
)(z)(\phi^i\c1\theta_k\wedge\c2 \phi^j\theta_l)(0)} + \oint zdz\wick{(\c1\rho_z^m\partial_z\c2{\bar \phi}_m)(z)(\c2\phi^i\theta_k\wedge \phi^j\c1\theta_l)(0)}\\
&=\delta^j_k\phi^i\theta_l -(-1)^{|^i_k|\cdot |^j_l|}\delta^i_l\phi^j\theta_k\\
&=\Big(\delta^j_k\delta^i_m\delta^n_l - (-1)^{|^i_k|\cdot |^j_l|}\delta^i_l\delta^j_m\delta^n_k\Big)\phi^m\theta_n\\
&=f_{^{ij}_{kl}}^{\ \ \ ^m_n} \phi^m\theta_n.
\end{split}
\end{align}
Then, one obtains the first constraint:
\begin{equation}
B^{ji}_{kl}\ f_{^{ji}_{kl}}^{\ \ \ ^m_n} =  0.
\end{equation}
\begin{itemize}
\item $b_0\Big(B^a_{\ \hat a}U_a^{\ \hat a}\Big)=0$:
\end{itemize}
\begin{align}
\begin{split}
&\ b_0(\theta_a\wedge\phi^{\hat a}(\phi\cdot\theta)) = \oint zdz\wick{(\c1\rho_z^m\partial_z\c2{\bar \phi}_m)(z)(\theta_a\wedge\phi^{\hat a}(\phi\cdot\theta))(0)}\\
=&\oint zdz\wick{(\c1\rho_z^m\partial_z\c2{\bar \phi}_m)(z)(\c1\theta_a\wedge\phi^{\hat a}(\c2\phi\cdot\theta))(0)} + \oint zdz\wick{(\c1\rho_z^m\partial_z\c2{\bar \phi}_m)(z)(\c1\theta_a\wedge\c2\phi^{\hat a}(\phi\cdot\theta))(0)}\\
=&\phi^{\hat a}\theta_a + (-1)^{|a|\cdot|\hat a|}\delta_a^{\hat a}(\phi\cdot\theta)\\
=&\Big(\delta^{\hat a}_m\delta^n_a + (-1)^{|a|\cdot|\hat a|} \delta^{\hat a}_a\delta^n_m\Big)\phi^m\theta_n\\
=&f_{a\hat a}^{\ \ ^m_n}\phi^m\theta_n.
\end{split}
\end{align}
Then, the one obtains the second constraint:
\begin{equation}
B_a^{\ \hat a}f_{a\hat a}^{\ \ ^m_n} = 0.
\end{equation}

\begin{itemize}
    \item $b_0\Big(B_i^{ja}U^i_{ja}\Big)=0$:
\end{itemize}
\begin{align}
\begin{split}
b_0(\theta_a\wedge\phi^i\theta_j) &=  \oint zdz\wick{(\c1\rho_z^m\partial_z\c2{\bar \phi}_m)(z)(\c1\theta_a\wedge\c2\phi^i\theta_j)(0)}\\
&= \delta^i_a\theta_j = \delta^i_a\delta_j^b\theta_b= f_{a\ ^i_j}^{\ \ \ b}\theta_b.
\end{split}
\end{align}
Following the third constraint:
\begin{equation}
B^{ai}_{\ j}f_{a\ ^i_j}^{\ \ \ b} = 0.
\end{equation}
\begin{itemize}
\item  $b_0\Big(B_{i\hat a}^jU_j^{i\hat a}\Big)=0$:
\end{itemize}
\begin{align}
&\ b_0(\phi^{\hat a}(\phi\cdot\theta)\wedge\phi^i\theta_j) =  \oint zdz\wick{(\c1\rho_z^m\partial_z\c2{\bar \phi}_m)(z)(\phi^{\hat a}(\phi\cdot\theta)\wedge\phi^i\theta_j)(0)}\nonumber\\
&= \oint zdz\wick{(\c1\rho_z^m\partial_z\c2{\bar \phi}_m)(z)(\phi^{\hat a}(\phi\cdot\c1\theta)\wedge\c2\phi^i\theta_j)(0)} + \oint zdz\wick{(\c1\rho_z^m\partial_z\c2{\bar \phi}_m)(z)(\phi^{\hat a}(\c2\phi\cdot\theta)\wedge\phi^i\c1\theta_j)(0)}\nonumber\\
&\hspace{1cm} + \oint zdz\wick{(\c1\rho_z^m\partial_z\c2{\bar \phi}_m)(z)(\c2\phi^{\hat a}(\phi\cdot\theta)\wedge\phi^i\c1\theta_j)(0)}\\
&=\phi^{\hat a}\phi^i\theta_j - \phi^{\hat a}\phi^i\theta_j + \delta^{\hat a}_j\phi^i(\phi\cdot\theta)=\delta^{\hat a}_j\phi^i(\phi\cdot\theta) = \delta^{\hat a}_j\delta^i_{\hat b}\phi^{\hat b}(\phi\cdot\theta) = f_{^i_j\ \hat a}^{\ \ \ \hat b}\phi^{\hat b}(\phi\cdot\theta).\nonumber
\end{align}
And then, the last constraint is:
\begin{equation}
B^{^i_j\ \hat a}f_{^i_j\ \hat a}^{\ \ \ \hat b} = 0.
\end{equation}

To summarize this computation, we can schematically explain what is happening by first writing the generators in (\ref{localghost2}) as:
\begin{equation}
U_{AB} = t_A\wedge t_B,
\end{equation}
where $t_A = (t_a,t^{\hat a},t^i_j)$. The operators in the $\mathfrak{psl}\wedge\mathfrak{psl}$ multiplet of (\ref{oplus}) are characterized by the $t_A$ being super-traceless operators. By the computation above, the gauge fixing implies that:
\begin{align}
\begin{split}
& b_0\Big( B^{AB}t_A\wedge t_B\Big) = B^{AB}[t_A,t_B] = B^{AB}f_{AB}^{\ \ \ C}t_C = 0.
\end{split}
\end{align}

What these constraints are saying is that the $\mathfrak{psl}$-components, $t_A$, of the operator have a vanishing internal commutator. Therefore, the operators lies in the irreducible representation of the $PSL(4|4)$ symmetry algebra, described in (\ref{def:(g^g)0}):
\begin{equation}
V_2\in(\mathfrak{psl}\wedge\mathfrak{psl})_0,
\end{equation}
which characterizes the beta deformation states.

As explained in the section \ref{betadeformation}, the beta deformation states are identified as operators living in the following irreducible representation of the symmetry algebra:
\begin{equation}\label{covering}
\frac{(\mathfrak{psl}\wedge\mathfrak{psl})_0}{\mathfrak{psl}}.
\end{equation}

This means that the beta deformation multiplets of the twistor string are described by the following vertices:
\begin{align}\label{betadefvert}
\begin{split}
V_2 = &\ \ B^{ij}_{kl}\phi^k\phi^l\theta_i\theta_j + B^{ia}_{j}\phi^j\theta_i\theta_a + B^{i}_{j\hat a}\phi^j\phi^{\hat a}\phi^k\theta_i\theta_{k}\\
& + B_{\hat a\hat b}\phi^{\hat a}\phi^i\phi^{\hat b}\phi^j\theta_i\theta_j + B^{\ a}_{\hat a}\phi^i\phi^{\hat a}\theta_i\theta_a + B^{ab}\theta_a\theta_b,
\end{split}
\end{align}
where $B$ are constant tensors satisfying the conditions to be in the irreducible representation (\ref{covering}), which can be translated by the constraints (\ref{tracelessB}) and (\ref{vanishinginternalcommutator}) found during this section, which we summarize now:
\begin{enumerate}
\item Anti-symmetric condition: This is implied by the cohomology computation and imposes that $V_2\in\mathfrak{pgl}\wedge\mathfrak{pgl}$;
\item Traceless condition:
\begin{equation}
B^{ij}_{il} = B^{ij}_{kj} = B^{ia}_{i} = B^{i}_{i\hat a} = 0.
\end{equation}
This imposes that the operator lives on the following multiplet of (\ref{oplus}): $V_2\in\mathfrak{psl}\wedge\mathfrak{psl}$;
\item Internal commutator condition:
\begin{equation}
B^{ji}_{kl}\ f_{^{ji} _{kl}}^{\ \ \ ^m_n}  = B_a^{\ \hat a}f_{a\hat a}^{\ \ ^m_n} = B^{i\ a}_{j}f_{^i_j\ a}^{\ \ \ b} = B_j^{i\ \hat a}f_{^i_j\ \hat a}^{\ \ \ \hat b} = 0.
\end{equation}
This condition is due to the gauge fixing, and imposes that $V_2\in(\mathfrak{psl}\wedge\mathfrak{psl})_0$;
\item Equivalence relation: This is given by relation (\ref{equiv.rel}), and is also imposed by the cohomological computation. It implies the final conclusion that the operator lives in the irreducible representation that characterizes beta-deformation: $[V_2]\in\frac{(\mathfrak{psl}\wedge\mathfrak{psl})_0}{\mathfrak{psl}}$. Indeed, the vertex operator $V_2$ in equation (\ref{V2}) is a representative of the equivalence class in which the operator lives. The local coordinates projected the equivalence class into this representative.

\end{enumerate}

\subsubsection{Reality condition vs Chirality}

We have finally described the beta deformation sates of the twistor string. Now, we will analyze this result and conclude that this description implicitly assumes a reality condition. Moreover, we will see that without this reality condition, the beta deformation states of the twistor string carries a certain chirality, as we explain next.

Consider the beta deformation in local coordinates, given by (\ref{ghost2}). It is parametrized by the tensor $B^{{\cal I}{\cal J}}_{{\cal K}{\cal L}}$, which is antisymmetric in ${\cal I}{\cal J}$ and symmetric in ${\cal K}{\cal L}$. The general beta deformation described in \cite{andrei-rivelles} is given by a sum of two tensors, one with antisymmetric in lower indices and symmetric in upper indices, and other with symmetric in lower indices and antisymmetric in upper indices\footnote{Here, since we are dealing with super-algebras, the symmetrization takes into account the statistics:
\begin{align}
\begin{split}
&A^{({\cal I}{\cal J})} = A^{{\cal I}{\cal J}} + (-1)^{{|{\cal I}||{\cal J}|}} A^{{\cal J}{\cal I}}\\
&A^{[{\cal I}{\cal J}]} = A^{{\cal I}{\cal J}} - (-1)^{{|{\cal I}||{\cal J}|}} A^{{\cal J}{\cal I}}
\end{split}
\end{align}} :
\begin{equation}\label{sym+asym}
B^{{\cal I}{\cal J}}_{{\cal K}{\cal L}} = B^{({\cal I}{\cal J})}_{[{\cal K}{\cal L}]} + B^{[{\cal I}{\cal J}]}_{({\cal K}{\cal L})}.
\end{equation}
Then, in principle, the topological string contains only half of the beta-deformation states. It turns out, however, that the two tensors (\ref{sym+asym}) can be related by a reality condition:
\begin{equation}\label{reality}
B^{({\cal I}{\cal J})}_{[{\cal K}{\cal L}]} =  \left(B^{[{\cal K}{\cal L}]}_{({\cal I}{\cal J})}\right)^*.
\end{equation}

Thenrefore, the description of the beta deformation is given by the $4$ conditions listed at the end of the last subsection, together with the reality condition (\ref{reality}). So far, we have implicitly assumed this reality condition, which allows us to characterize the ghost number two operators as ${\frak{g}\wedge\frak{g}\over\frak{g}}$.

From this analysis, we conclude that in the general case where we do not assume the reality condition (\ref{reality}), the beta deformation of twistor string is \textit{chiral}, in the sense that it only contains half of the beta-deformation states (\ref{sym+asym}).

\section{Deformation of the action}

In the last sections, we constructed and studied a ghost number two operator that corresponds to the beta-deformation. However, we still don't know how this operator deforms the topological action. In this section, we will investigate this question, by describing a new topological action, obtained when the vertex operator deforms the original action. To address this question, one needs to first understand the difference between an integrated and unintegrated vertex operator.

The operator (\ref{betadefvert}) is an unintegrated vertice, since it is a zero form on the worldsheet. Let us call it $V_\beta^{(0)}$. Its correspondent integrated vertices, $V_\beta^{(1)}$ and $V_\beta^{(2)}$, can be obtained by the following descent equations:
\begin{align}\label{descent}
\begin{split}
&dV_\beta^{(0)} = QV_\beta^{(1)},\\
&dV_\beta^{(1)} = QV_\beta^{(2)}.
\end{split}
\end{align}

These equations holds because the topological invariance of the theory implies that the correlation functions of $V_\beta(z)$ are independent of the point in the worldsheet $z$. Therefore, it follows that the operators $V^{i}_\beta$ must be $d$-closed, up to a BRST-exact term $QV_\beta^{(i+1)}$ \cite{WittenTop}. The operator $V_\beta^{(1)}$ is a 1-form on the worldsheet and have ghost number 1. If $C$ is a circle in the worldsheet, then, we can define a new BRST invariant physical observable:
\begin{equation}
V(C)=\oint_C V_\beta^{(1)}.
\end{equation}

The operator $V_\beta^{(2)}$ is a 2-form on the worldsheet with ghost number zero. Again, we can define a new BRST invariant physical observable by integrating it over the worldsheet:
\begin{equation}
W_\beta=\int_\Sigma V_\beta^{(2)}.
\end{equation}
The operator $W_\beta$ leads to a new topological sigma model, that defines the beta-deformed twisted string:
\begin{equation}\label{deformedaction}
S_\beta = S_0 + \int_\Sigma V_\beta^{(2)},
\end{equation}
where $S_0$ is the original lagrangian (\ref{model}). We now turn to the explicit computation of this deformation. Let us first write $V_\beta^{(0)}$ as:
\begin{align}\label{vertexzero}
&V^{(0)}_\beta = V_\beta^{ij}(\phi)\theta_i\theta_j,\ \ \ \ \text{where:}\\
&V_\beta^{ij}(\phi) = B^{ij}_{kl}\phi^k\phi^l + B^{ia}_{j}\delta_a^j\phi^j + B^{i}_{k\hat a}\phi^k\phi^{\hat a}\phi^j+ B_{\hat a\hat b}\phi^{\hat a}\phi^i\phi^{\hat b}\phi^j + B^{\ a}_{\hat a}\delta_a^j\phi^i\phi^{\hat a} + B^{ab}\delta^i_a\delta^j_b.\nonumber
\end{align}

The descent equations in (\ref{descent}) is valid only up to equations of motion. This fact will lead to a deformation of the BRST operator of the model. The first descent equation is:
\begin{equation}
dV_\beta^{(0)} = QV_\beta^{(1)} + 2V_\beta^{ij}(\phi)\theta_i\frac{\delta L}{\delta\rho^j},
\end{equation}
with 
\begin{align}\label{V1}
\begin{split}
& V_\beta^{(1)} = \rho^k\partial_kV_\beta^{ij}(\phi)\theta_i\theta_j - 2V_\beta^{ij}(\phi)\theta_i\star d\bar\phi_{j},
\end{split}
\end{align}
where $\star$ is the Hodge star operator. The second descent equation is:
\begin{align}\label{descent2}
\begin{split}
&dV_\beta^{(1)} = QV_\beta^{(2)} +  \frac{\delta L}{\delta\phi^i}\zeta^i + \frac{\delta L}{\delta\theta^i}\xi^i,\\
&\text{where:} \ \ \ \zeta^i = -2V_\beta^{ij}(\phi)\theta_j\\
&\hspace{1cm}\ \ \ \ \xi^i = \partial_iV_\beta^{kl}(\phi)\theta_k\theta_l.
\end{split}
\end{align}

Equation (\ref{descent2}) means the new action (\ref{deformedaction}) is no longer invariant under the original BRST operator $Q$, but is invariant under a new, deformed BRST operator:
\begin{equation}\label{BRSTdeformed}
Q^{(\beta)} = \eta^{\bar i}\frac{\partial}{\partial\bar\phi^{\bar i}} + d\phi^i\frac{\partial}{\partial\rho^i}  -2V_\beta^{ij}(\phi)\theta_j\frac{\delta}{\delta\phi^i} + \partial_iV_\beta^{kl}(\phi)\theta_k\theta_l\frac{\delta}{\delta\theta^i}.
\end{equation}

The structure of the deformation of the BRST operator is the same as that found in the pure spinor case \cite{andrei-rivelles}, which takes the form
\begin{equation}
\delta Q = B^{AB} \Lambda_A t_B,
\end{equation}
where \( t_A \) are the generators of the superconformal algebra, and \( \Lambda \) satisfies \( Q \Lambda = dj \), with \( j \) the superconformal conserved current. Moreover, if we identify the initial BRST operator with the \( \bar\partial \) operator, we see that equation (\ref{BRSTdeformed}) is a perturbation of $\bar\partial$ given by:
\begin{equation}\label{Qbeta}
Q^{\beta} = \bar{\partial} + [V^{(0)}_\beta, -],
\end{equation}
where $[\cdot, \cdot]$ denotes the Schouten bracket.

One can see that the deformed BRST operator $Q^{(\beta)}$ is nilpotent when the matrix $B$ satisfies the Classical Yang-Baxter equation (cYBe):
\begin{equation}\label{YBe}
[BX,BY] - B[X,BY] - B[BX,Y] = 0,
\end{equation}
for general matrices $X,Y$. The YB equation above can also be written as in the form:
\begin{equation}
[\![ B,B ]\!]^{abc} = B^{[a|d}f_{de}{}^bB^{e|c]} = 0.
\end{equation}

An important observation is that we can consider a more general matrix $B$, and this would also give us interesting models. For instance, as we will comment in the next section, $B$ could satisfies the modified Classical Yang-Baxter equation (mcYBe):
\begin{equation}\label{mYBe}
[BX,BY] - B[X,BY] - B[BX,Y] + c^2[X,Y] = 0,
\end{equation}
for $c=1,i$. In this case, since the BRST operator $Q^\beta$ in (\ref{Qbeta}) is no longer nillpotent, it should receive further corrections, of quadratic order in the deformation.

\subsection{Beta deformation as a current-current deformation}

The expression for the operator $V^{(2)}_\beta$ is more complicated to find. A much more simple way to obtain it is by the use of the superspace formulation, which was used in in this context in \cite{kodaiara-spencer}. Using this approach, we can resolve the descent equations by using the composite $b$-ghost. One can solve the equations (\ref{descent}) as follows:
\begin{align}\label{solutiondescent}
\begin{split}
&V_\beta^{(1)} = \{b-\bar b,V_\beta^{(0)}\},\\
&V_\beta^{(2)} = \{b,[\bar b,V_\beta^{(0)}]\},
\end{split}
\end{align}
where the brackets means the single pole from the OPE. We can see that the first equation above agrees with the computation for $V^{(1)}_\beta$ in (\ref{V1}). Now, using the second equation, we will be able to characterize the deformation of the action as a product of two conserved currents of the $PSL(4|4)-$symmetry:
\begin{align}\label{integratedbeta}
\begin{split}
V_\beta^{(2)} = &\ \ B^{ij}_{kl}\ J^k_i\wedge J^l_j + B^{ia}_{j}\ J^j_i\wedge J_a + B^{i}_{j\hat a}\ J^j_i\wedge J^{\hat a}\\
& + B_{\hat a\hat b}\ J^{\hat a}\wedge J^{\hat b} + B^{\ a}_{\hat a}\ J^{\hat a}\wedge J_a + B^{ab}\ J_a\wedge J_b,
\end{split}
\end{align}
where $J$ are the $\mathfrak{psl}-$conserved currents, which reads as:
\begin{align}
\begin{split}
& J^i_j = \phi^i\star d\bar\phi_j + \rho^i\theta_j\ +\ \text{c.c}\ ,\ \ J_a = \star d\bar\phi_a\ +\ \text{c.c},\\
& \ \ \ \ J^{\hat a} = \phi^{\hat a}\phi^i\star d\bar\phi_i + \rho^{\hat a}\phi^i\theta_i + \phi^{\hat a}\rho^i\theta_i\ +\ \text{c.c}.
\end{split}
\end{align}

\section{Applications to integrable deformations}

We conclude this paper with a discussion of the further application of our results. A key observation is that the deformation of the topological string we have found corresponds to a current-current deformation (\ref{integratedbeta}). This type of deformation is well studied in the context of integrable deformations \cite{borsato-jj,Hoare:notes,Ben-Ryan,Ben-Ryan:diamond}, suggesting the possibility of an underlying integrability structure within the twistor string model explored in this work.

Further evidence supporting the presence of such a structure is that the constant matrix $B$ is antisymmetric and satisfies the Yang-Baxter equation (\ref{mYBe}). This class of matrices plays a fundamental role in integrable deformations, giving rise to Yang-Baxter models \cite{Hoare:notes}. While a full investigation of this aspect is beyond the scope of this paper, we will outline a framework for bridging our results to a more detailed study of integrability in this setting.

The simplest integrable deformation is the Leigh-Strassler deformation \cite{Leigh-Strassler} of the ${\cal N}=4$ SYM, which corresponds to an integrable deformation of the strings in $AdS_5\times S^5$, usually called abelian Yang-Baxter deformation \cite{borsato-jj}. The work developed in \cite{Kulaxizi:2004} reproduces the Leigh-Strassler deformation from the perspective of the holomorphic Chern-Simons (hCS) theory, which describes the open string sector of the B-model. There, the hCS is deformed with a star product. In this work, we deform the action of the B-model with a corresponding operator, that we will write next.

It is important to note that in this work we take a more systematic and comprehensive approach to understanding deformations. Specifically, we addressed the problem of classifying deformations by identifying a complete multiplet of states corresponding to a broad class of deformations, which we refer to generally as beta-deformations. This classification builds on the framework introduced in \cite{andrei-rivelles} in the context of the pure spinor superstring in $AdS_5 \times S^5$.

We can identify a subset of deformations that correspond to the Leigh-Strassler deformations. Since this deformation breaks $\mathcal{N}=4$ supersymmetry down to $\mathcal{N}=1$, we will express our results in an $SU(3) \times U(1)$-invariant form. To do so, we decompose the fermionic coordinates $\psi^A$ of $\mathbb{CP}^{3|4}$ as $\psi^A = \{\xi, \psi^I\}$, where $I = 1,2,3$. Next, we consider the vertex operator that carries both indices in the fermionic directions, specifically in the components $\psi^I$ associated with the $SU(3)$ sector. Using global coordinates, the vertex operator corresponding to the Leigh-Strassler deformations is given by
\begin{equation}
V = B^{IJ}_{KL}\psi^K\psi^L\Theta_I\Theta_J,
\end{equation}
where the constant matrix $B$ is explicitly given by:
\begin{equation}
B^{IJ}_{KL} = h^{IJN}\epsilon_{NKL},
\end{equation}
where $h^{IJN}$ and $\epsilon_{NKL}$ are totally anti-symmetric tensors with $\epsilon_{123}=1$ and $h^{123}=\beta$, where $\beta$ is the deformation parameter. The 3-parameter deformation, considered in \cite{Fokken:gammai}, follows the same construction, but now we should take the tensor $h$ to be of the form:
\begin{equation}
h^{123} = -h^{132} = \gamma_1,\ \ \ h^{231} = -h^{213} = \gamma_2,\ \ \ h^{312} = -h^{321} = \gamma_3.
\end{equation}

In \cite{Benoit:letter,q-deformation}, a new interesting class of Yang-Baxter deformations was considered, with huge applications to superstrings in $AdS_5\times S^5$. There, what characterizes the deformations is an $R$-matrix that satisfies the modified classical Yang-Baxter equation (\ref{mYBe}), again:
\begin{equation}
[RX,RY] - R[X,RY] - R[RX,Y] + c^2[X,Y] = 0.
\end{equation}
for $c=i,1$. We claim that we can reproduce these deformations in the context of the B-model, if we specify our matrix $B$ to satisfy the mcYBe above. In this case, the BRST operator (\ref{Qbeta}) receive further corrections, when we require it to be nillpotent.

\subsection{Non-commutative spaces}

One of the implications of our work is that it encompasses deformations that give rise to non-commutative Yang-Mills theories \cite{vanTongeren, twist-noncommutative, Meier}. These deformations break the conformal structure but remain of significant interest due to the emergence of a new class of symmetries, known as twistor symmetries.

The Leigh-Strassler deformation preserves conformal symmetry, as it corresponds to a deformation along the R-symmetry directions. However, whenever the vertex operator transforms under the subalgebra of $PSU(2,2|4)$ corresponding to the ordinary conformal algebra $SU(2,2)$, the conformal invariance of the theory is broken. The distinction between deformations involving the R-symmetry and those involving the conformal algebra can be seen by examining the block matrix in equation (B.14).

We then can write explicitly the vertex operator corresponding to the non-commutative Yang-Mills theories. We first write the coordinates of $\mathbb{CP}^{3|4}$ as $\Phi=\{\phi^\mu,\psi^A\}$, and then the vertex operator we want is:
\begin{equation}
V = B^{\mu\nu}_{\rho\sigma}\phi^\rho\phi^\sigma\Theta_\mu\Theta_\nu.
\end{equation}
This vertex operator induces a star (non-commutative) product on the superspace, which is obtained from the twist:
\begin{equation}\label{twist}
{\cal F} = \exp\left( \frac{1}{2}B^{\mu\nu}_{\rho\sigma}\phi^\rho\partial_\mu\wedge\phi^\sigma\partial_\nu \right).
\end{equation}
The corresponding star-product is
\begin{equation}\label{starproduct}
{\cal A}(\phi)\star {\cal B}(\phi) \equiv {\cal A}(\phi){\cal B}(\phi) + \sum_{n=1}^\infty \frac{1}{2^nn!}\left(B^{\mu\nu}_{\rho\sigma}\phi^\rho\phi^\sigma\partial_\mu\partial'_\nu\right)^n {\cal A}(\phi){\cal B}(\phi')\Big|_{\phi=\phi'}.
\end{equation}

In order to stablish the direct connection with \cite{twist-noncommutative,Meier}, it is usefull to write the bosonic coordinates of $\mathbb{CP}^{3|4}$ as $\phi=(\lambda^\alpha;\mu^{\dot\alpha})$. With this notation, we can explicitly identify the generators of the Poincaré algebra, as given in (\ref{t})–(\ref{d}). In this case, the Groenewold-Moyal noncommutative space is obtained from a twist that involves translations:
\begin{equation}
{\cal F} = \exp\left( \frac{1}{2}B^{\dot\alpha\dot\beta}_{\alpha\beta}\lambda^\alpha\partial_{\dot\alpha}\wedge\lambda^\beta\partial_{\dot\beta} \right),
\end{equation}
where $\partial_{\dot\alpha} = \frac{\partial}{\partial\mu^{\dot\alpha}}$. On the other hand, the Quadratic twist-noncommutative gauge theory, thoroughly studied in \cite{Meier}, can be obtained from the twist involving rotations:
\begin{equation}
{\cal F} = \exp\left( \frac{1}{2}B^{\alpha\dot\alpha}_{\beta\dot\beta}j^\alpha_\beta\wedge j^{\dot\alpha}_{\dot\beta}\right),
\end{equation}
where $j^\alpha_\beta = \frac{i}{2}\left( \lambda_\alpha\frac{\partial}{\partial\lambda^\beta} + \lambda_\beta\frac{\partial}{\partial\lambda^\alpha}\right)$ and $j^{\dot\alpha}_{\dot\beta} = \frac{i}{2}\left( \mu_{\dot\alpha}\frac{\partial}{\partial\mu^{\dot\beta}}+ \mu_{\dot\beta}\frac{\partial}{\partial\mu^{\dot\alpha}}\right)$. Finally, the Lie-algebraic noncommutative space mixes translations and rotations:
\begin{equation}
{\cal F} = \exp\left( \frac{1}{2}B^{\dot\alpha\dot\beta}_{\alpha\dot\gamma}\lambda^\alpha\partial_{\dot\alpha}\wedge j^{\dot\gamma}_{\dot\beta}\right),\ \ \ \ \ {\cal F} = \exp\left( \frac{1}{2}B^{\dot\alpha\beta}_{\alpha\gamma}\lambda^\alpha\partial_{\dot\alpha}\wedge j^{\gamma}_{\beta}\right),
\end{equation}

One may also ask for the other twists included in (\ref{twist}), but not considered in \cite{twist-noncommutative}. For instance the ones involving the Special Conformal Transformation (SCT) generators.

Moreover, one could also consider the theory coming from a twist that mixes the generators of the conformal algebra with the generators of the R-symmetry:
\begin{equation}
{\cal F} = \exp\left(\frac{1}{2}B_{IA}^{JB}\phi^I\partial_J\wedge\psi^A\partial_B\right),
\end{equation}
where $\partial_I=\frac{\partial}{\partial\phi^I}$ and $\partial_A=\frac{\partial}{\partial\psi^A}$. Similarly, one can consider the twists involving fermionic generators:
\begin{equation}
{\cal F} = \exp\left(\frac{1}{2}B_{IA}^{BJ}\phi^I\partial_B\wedge\psi^A\partial_J\right).
\end{equation}

\subsection{Open string sector: holomorphic Chern-Simons}

It wight be interesting to investigate the open string sector of the B-model, and then connects the present work with the integrable deformations obtained from the holomorphic Chern-Simons, which is the string field theory of the topological B-model, as studied in \cite{Ben-Ryan}. Moreover, the connection between this approach and the pure spinor formalism was established in \cite{rodrigo}. In this sense, understanding how the beta deformation in the pure spinor formalism comes from the four-dimensional Chern-Simons theory could provide valuable insights.

In \cite{Kulaxizi:2004}, the open sector of the B-model is deformed with a star product. In this work, we deformed the B-model itself. One can then try to look to the open string sector of our deformed model, and then try to relate it to the deformed open sector in \cite{Kulaxizi:2004}. The open string sector of our deformed model can be formulated by replacing $\bar\partial$ of the undeformed model, by the new BRST operator $Q^\beta$ in (\ref{Qbeta}). The hCS would then take the form of:
\begin{equation}
S = \int_{Y} \Omega\wedge \tr\left({\cal A} Q^{\beta}{\cal A} + \frac{2}{3}{\cal A}\wedge{\cal A}\wedge{\cal A}\right),
\end{equation}
where ${\cal A}$ encodes the SYM superfields, $\Omega$ is the $Q^\beta$-holomorphic $(3,0)$-form of the Calabi–Yau manifold, and the $Y$ is the world-volume of D-branes, identified with the submanifold $\bar\psi^{\bar A}=0$ within $\mathbb{CP}^{3|4}$, as described in \cite{twistor,Kulaxizi:2004}. In this way, we establish a connection between the open sector of our deformed theory and the deformed open sector of the undeformed theory considered in \cite{Kulaxizi:2004}. In the amplitude computation, to each incoming field we associate a wavefunction $w_i$. This wavefunction is $\bar\partial$-closed in the undeformed B-model. In our case, this wavefunction is no longer $\bar\partial$-closed, but $Q^\beta$-closed. Moreover, the $(3,0)$-form $\Omega$ is also deformed in order to also be $Q^\beta$-closed. Although still not completely clear, this correction to the $\bar\partial$ operator should be equivalent to the $\star$-product. If this is true, the open sector of the deformed theory is equivalent to the deformed open sector studied in \cite{Kulaxizi:2004}. The introduction of a star product gives us a non-local theory in a non-commutative superspace. One of the advantages of our approach is that we just modify the BRST operator and still have a local theory in a regular superspace.

One of the possible immediate application of formulating the deformations in terms of a Chern-Simons theory is that it allows the computation of tree-level amplitudes, as was done in the undeformed theory in \cite{twistor} and in the LS deformation in \cite{Kulaxizi:2004}. This can potentially be used to obtain the amplitudes of the theories described in \cite{twist-noncommutative,Meier}.

\section{Discussions and future directions}

In this work, we described the beta deformation of the twistor string, which is formulated by the topological B-model on $\mathbb{CP}^{3|4}$. This was achieved by computing the cohomology of the BRST operator associated with the topological model. The beta deformation was identified with the states living in the irreducible representation (\ref{rep_betadef}) of the superconformal algebra. To identify these states, we applied the Siegel gauge to the ghost number two operators found from the cohomology.

In Section (\ref{ghostnumber1}), we showed that the ghost number one operators can be interpreted as the generators of symmetries of the $PSL(4|4)$ supergroup. An interesting observation arises when we apply equation (\ref{solutiondescent}) to the ghost number one operators: we obtain the conserved currents of the $PSL(4|4)$ supergroup. Although this result was expected—since the generators and currents must contain the same information—it serves as an important check on the procedure developed in this paper.

We also observed that the correction (\ref{BRSTdeformed}) to the BRST operator acting on $\phi$ takes the form $\delta Q = B^{AB} \Lambda_A t_B$. Here, $\Lambda_A$ represents ghost number one operators, which, by the descent equations explained above, satisfy the relation $d\Lambda_A = Q j_A$; while $t_A$ denotes a $\mathfrak{psl}$-symmetry vector field. The structure of this correction is the same as that found in the pure spinor case \cite{andrei-rivelles}. One may now ask whether this structure also leads to some integrability properties of the deformed model.

Another related topic concerns topological dualities. The concept of \textit{twisted holography} was recently introduced in \cite{TwistedHolography} as a duality that connects the B-model topological string theory with two-dimensional chiral algebras. A natural question arising from this framework is how we can interpret the beta-deformation discussed in this paper from the perspective of the chiral algebras that are dual to the B-model.

The study of deformations in twisted $N=2$ theories was thoroughly explored in \cite{kodaiara-spencer}. The general deformation of these models is of the following form:
\begin{equation}
S = S_0 + t_i\int_\Sigma V_i,
\end{equation}
where $V_i$ are elements of the BRST cohomology of the topological model, parametrized by $t_i$. These operators must have conformal weight two. The effect of this deformation leads to an anomaly in the partition function at every genus, known as the holomorphic anomaly. This is captured to all orders by a \textit{master anomaly equation}, known as the BCOV equation. One should try to understand what is the efect of the operators we computed in this paper, summarized in equation (\ref{vertexzero}), in this context of holomorphic anomalies.

Finally, it would be worthwhile to revisit the problem discussed here from a mathematically rigorous perspective, particularly through the lens of Homological Mirror Symmetry, as developed in \cite{Veronica:thesis,veronica:advancements}. In this framework, one could analyze the deformations considered in this work in terms of the categorical structures introduced in these references. Moreover, it would be interesting to explore how these deformations manifest in the A-model, which is mirror dual to the B-model considered here. Such an approach could provide further insight and shed new light on the discussion presented in this work.

\section*{Acknowledgments}

This work was supported in part by ICTP-SAIFR FAPESP grant 2019/21281-4 and by FAPESP grant 2022/00940-2. I would like to thank Andrei Mikhailov for many enlightened discussions and for proposing ideas for this work. I also want to thank Horațiu Năstase, Marcelo Barbosa, Tim Meier, Veronica Pasquarella for usefull discussions, and S. Noja for the help with the cohomological computation on projective space. Finally, I want to thank Thiago Fleury for the comments on this draft.

\appendix

\section{Projective space}\label{ApxProjective}

The twistor string will be defined on a projective space of super-dimension $3|4$. In general, the projective space is defined as the space of the following equivalence classes:
\begin{align}\label{cp}
&\mathbb{CP}^{n|m} \equiv \mathbb{C}^{n+1|m}\Big/\sim\\
&\text{where}\ \ \   x \sim t\cdot x,\ \ \forall t\in\mathbb{C}^*.\nonumber 
\end{align}
Here we define the super-space $\mathbb{C}^{n|m}$ as a supercommutative free module:
\begin{equation}
\mathbb{C}^{n|m} \equiv \mathbb{C}^{n}\oplus\Pi\mathbb{C}^{m},
\end{equation}
where $\Pi$ is the parity changing functor that indicates the reversing of the parity. We can define the projection map:
\begin{align}
\pi:&\ \mathbb{C}^{n+1|m}\longrightarrow\mathbb{CP}^{n|m}\\
&\ \ X\longmapsto [X]:=\{Y\in\mathbb{C}^{n+1|m}\ |\ Y\sim X\}.
\end{align}

We can define charts in the projective space $\mathbb{CP}^{n|m}$ given by $\pi(U_j)$, where $U_j$ are open sets in $\mathbb{C}^{n+1|m}$ defined by:
\begin{equation}
U_j := \{ X\in\mathbb{C}^{n+1|m}\ |\ X_j\ne0 \},\ \ \ {j=0,\cdots,n}.
\end{equation}
The charts are defined as a pair of an open set together with a function from this open set to an euclidean space: $(\pi(U_j),\varphi_j)$. As an illustrative example, the map $\varphi_0$ is defined by:
\begin{align}\label{chart}
\varphi_0:  \pi(U_0) & \subset \mathbb{CP}^{n|m} \longrightarrow \mathbb{C}^{n|m}\nonumber\\
&[X] \longrightarrow \Vec{x} = \varphi_0([X]),\hspace{1cm} \text{where}\ x^j := \frac{X^j}{X^0},\ \text{for}\ j=1,\cdots,n+m.
\end{align}
where similarly for the other open sets $U_j$'s, we just need to properly organize the coordinates in the right hand side of (\ref{chart}). For instance, when $j=1$:
\begin{equation}
U_1\ni(X^0,X^1,\cdots,X^{n+m})\longrightarrow (x^0,x^2,\cdots,x^{n+m}),
\end{equation}
where $x^j := \frac{X^j}{X^1}$, for $j=0,2,\cdots,n+m$.

It is simple to see that the maps $\varphi_j$ are well defined. To that purpose, we have take two equivalent vectors $X\sim Y$ and see that they are mapped to the same vector. Under $\varphi_j$, the vector $X$ is mapped to $\Vec{x}$ with $x^i = \frac{X^i}{X^j}$. On the other hand, $Y$ is mapped to $\Vec{y}$ with $y^i = \frac{Y^i}{Y^j}$. However, since $X$ and $Y$ are equivalent vectors, these two are related by the equation $Y=tX$ for some $t$. Then, we conclude that $\Vec{x}=\Vec{y}$.

\subsection{Tangent bundle of projective space}\label{tangent-CP}

In order to describe the coordinates of the vector fields in the projective space, one chooses a chart $(U,\pi)$. Let us choose $U=U_0$, and the corresponding projection $\pi=\phi_0$. One then consider the image under $\pi$ of the tangent vector $\partial/\partial X^i$:
\begin{align}\label{d/dx}
\begin{split}
& \pi_*\frac{\partial}{\partial X_i} = \frac{1}{X_0}\frac{\partial}{\partial x_i},\\
& \pi_*\frac{\partial}{\partial X_0} = -\sum_i\frac{X_i}{X^2_0}\frac{\partial}{\partial x_i},\hspace{1cm} i=1,\cdots,n+m.
\end{split}
\end{align}
As well, one considers the image under $\pi$ of the 1-form $dX^i$:
\begin{align}\label{dx}
\begin{split}
& \pi_*dX_i = X_0dx_i,\\
& \pi_*dX_0 = -\sum_i\frac{X^2_0}{X_i}dx_i,\hspace{1cm} i=1,\cdots,n+m.
\end{split}
\end{align}
From this calculation, it follows that the basis of the tangent bundle scales as $\partial/\partial X^i\rightarrow t^{-1}\partial/\partial X^i$, while the basis of the cotangent bundle scales as $dX^i\rightarrow t dX^i$.

Consider a vector field $v$ and a 1-form $\omega$ defined on $\mathbb{C}^{n+1|m}$. $v$ and $\omega$ descends to the projective space $\mathbb{CP}^{n|m}$ when they are invariant under rescaling: $\pi_*v(tX) = \pi_*v(X)$ and $\pi_*\omega(tX) = \pi_*\omega(X)$ for any complex number $t$. In other words, if $v$ and $\omega$ are well defined in the projective space, they take the following form:
\begin{align}
v &= v^i(X)\frac{\partial}{\partial X_i},\ \ \ \ \ v^i(tX) = tv^i(X),\\
\omega &= \omega_i(X)dX^i,\ \ \ \ \ \omega_i(tX) = t^{-1}\omega_i(X).
\end{align}

It also follows from (\ref{d/dx}) and (\ref{dx}) the following relations:
\begin{align}
\sum_iX^i\frac{\partial}{\partial X^i} = \sum_i\frac{1}{X^i}dX^i = 0.
\end{align}

The tangent and cotangent bundle of the projective space are represented in local coordinates as:
\begin{align}
(X;V)&=(X^0,X^1,\cdots,X^{n+m};V^0,V^1,\cdots,V^{n+m}) \in TU,\\
(X;P)&=(X^0,X^1,\cdots,X^{n+m};P_0,P_1,\cdots,P_{n+m}) \in T^*U,
\end{align}
where $X^i$ parametrizes the points in the manifold, while $V^i$ and $P_i$ parametrizes points inside the fiber in a given point $X$, respectively $T_xU$ and $T^*_xU$. The fiber coordinates carries the equivalence relations $V\sim tV$ and $P\sim t^{-1}P$. The equivalence relation $X\sim tX$ implies a constraint on the bundles, which are given by:
\begin{equation}\label{constraintbundle}
g_{ij}X^iV^j = X^iP_i = 0,
\end{equation}
where $g_{ij}$ is the metric of the projective space.

The understanding of local coordinates of the cotangent bundle was really important in this paper, specially in section 5. Using the construction of this appendix, we can write the local coordinates as:
\begin{equation}
(X;P) \sim \left(1,\frac{X^1}{X^0},\cdots,\frac{X^n}{X^0};-X^iP_i,X^0P_1,\cdots,X^0P_n\right) \longmapsto (x^1,\cdots,x^n;p_1,\cdots,p_n) \in T^*U,
\end{equation}
where $x^i=X^i/X^0$ and $p_i = X^0P_i$ and we have used the constraint (\ref{constraintbundle}). One may back to global coordinates by the following map:
\begin{equation}
(x^1,\cdots,x^n;p_1,\cdots,p_n)\hookrightarrow (1,x^1,\cdots,x^n;-x^ip_i,p_1,\cdots,p_n)\sim \left(X^0,\cdots,X^n;P^0,\cdots,P^n\right)
\end{equation}
for $X^0$ arbitrary number, and $X^i:= X^0x^i$, $P_i:=p_i/X^0$.

\subsection{From Global coordinates to Local coordinates}\label{ApxProjective2}

Consider a set $\{\Phi^I\}$ of global coordinates in the projective space $\mathbb{CP}^{3|4}$. We consider the set of holomorphic global vector fields defined in this space. The basis to this set is:
\begin{equation}\label{tijGlobal}
t^I_J = \Phi^I\frac{\partial}{\partial \Phi^J}.
\end{equation}
We can see these elements as linear transformations on $\mathbb{C}^{4|4}$. From this point of view, a linear transformation can be written as an operator $A=A^J_I\Phi^I\frac{\partial}{\partial \Phi^J}$.

We want to describe $t^I_J$ in terms of local coordinates $\varphi^i=\Phi^i/\Phi^1$. This means the following replacement:
\begin{equation}
\vec\Phi \longmapsto  \left(\begin{array}{c}
    1 \\
    \ \\
   \vec \varphi \\
    \
\end{array}\right)
\end{equation}
Let us first re-write the components of the operator $A$ as:
\begin{align}
\begin{split}
&a\equiv A^1_1,\ \ \ b_j\equiv A^1_j,\ \ \ c^i\equiv A^i_1,\ \ \ m^i_j\equiv A^i_j,
\end{split}
\end{align}
with $i,j=2,\cdots,8$. The linear transformation $A$ in local coordinates reads:
\begin{equation}\label{confinlocal}
\left( \begin{array}{c|ccc}
   a\ \ \ & \ & \vec b & \ \\
   \midrule
   \ & \ & \ & \ \\
   \vec c\ \ \ & \ &  m^i_{\ j}\\
    \ & \ & \
\end{array}\right) \left(\begin{array}{c}
    1 \\
    \ \\
    \vec \varphi\\
    \
\end{array}\right) = \left(\begin{array}{c}
    a+\vec b\cdot \vec \varphi \\
    \ \\
    \vec c+m\cdot\vec \varphi\\
    \
\end{array}\right) \longmapsto  \left(\begin{array}{c}
    1 \\
    \ \\
    \frac{\vec c+m\cdot\vec \varphi}{a+\vec b\cdot \vec \varphi}\\
    \
\end{array}\right)
\end{equation}
Therefore, from (\ref{confinlocal}), we compute the infinitesimal generators to conclude that in the local coordinates $\{\phi_i\}$, we have the following basis to $\mathbb{C}^{31|32}$:
\begin{equation}
\Bigg\{\frac{\partial}{\partial\varphi^a},\ \ \ \varphi^{\hat a}\varphi^k\frac{\partial}{\partial\varphi^k},\ \ \ \varphi^i\frac{\partial}{\partial\varphi^j} \Bigg\} = \{t_a,\ t^{\hat a},\ t^i_j\}.
\end{equation}
Such base is constructed from the infinitesimal generator of the transformation $ f(\vec\varphi)=\frac{\vec c+m\cdot\vec \varphi}{a+\vec b\cdot \vec \varphi}$. We can consider $a=1$ and impose the condition for the trace $\Tr(m)=-1$. Follows the expansion:
\begin{equation}
\frac{c^i+m^i_j\varphi^j}{1+b_k\varphi^k} = c^i+m^i_j\varphi^j - b_km^i_j \varphi^j\varphi^k + \cdots .
\end{equation}
To get only the first order term, we put $m^i_j=\delta^i_j$ in the last term of the expansion, in such a way that the infinitesimal generator can be written as:
\begin{equation}
c^i\partial_i + m^i_j\varphi^j\partial_i - b_k\varphi^k\varphi^j\partial_j.
\end{equation}

The structure constants can be computed by considering the (anti-)commutators of the basis we just described:
\begin{align}
\begin{split}
&[t^i_j,t^k_l\}=f_{^{ik}_{jl}}^{\ \ ^m_n}t^m_n,\ \ \ [t_a,t_{\hat a}\} = f_{a\hat a}^{\ \ ^m_n}t^m_n,\ \ \ [ t_a,t^i_j \} = f_{a\ ^i_j}^{\ \ \ b}t_b, \ \ \ [ t_{\hat a},t^i_j \}=f_{\hat a\ ^i_j}^{\ \ \ \hat b}t_{\hat b},
\end{split}
\end{align}
where:

\begin{align}
\begin{split}
&f_{^{il}_{jl}}^{\ \ ^m_n} = \delta^k_j\delta_m^i\delta^n_l- (-1)^{|^i_j|\cdot |^k_l|}\delta^i_l\delta_m^k\delta^n_j\\
&f_{a\hat a}^{\ \ ^m_n} =  \delta^{\hat a}_m\delta^n_a + (-1)^{|a|\cdot|\hat a|} \delta^{\hat a}_a\delta^n_m\hspace{3cm} f_{\hat aa}^{\ \ ^m_n}=-(-1)^{|a|\cdot|\hat a|}f_{a\hat a}^{\ \ ^m_n}\\
&f_{a\ ^i_j}^{\ \ \ b} = \delta^i_a\delta^b_j\hspace{6.0cm} f_{^i_j\ a}^{\ \ \ b}= -(-1)^{|a|\cdot|^i_j|}f_{a\ ^i_j}^{\ \ \ b} \\
&f_{\hat a\ ^i_j}^{\ \ \ \hat b}= -\delta^{\hat a}_j\delta^i_{\hat b}\hspace{5.7cm} f_{^i_j\ \hat a}^{\ \ \ \hat b}= -(-1)^{|\hat a|\cdot|^i_j|}f_{\hat a\ ^i_j}^{\ \ \ \hat b}
\end{split}
\end{align}

\section{Super-conformal algebra}\label{A}

The super-conformal algebra with $\mathcal{N}=4$ supercharges is given by the finite dimensional lie algebra $\mathfrak{psu}(2,2|4)$, which is the real form of $\mathfrak{psl}(4|4)$. We can decompose this algebra in terms of semi-simple ones and a fermionic space:
\begin{equation}
\mathfrak{su}(2,2|4) = \mathfrak{su}(2,2)\oplus\mathfrak{su}(4)\oplus\mathfrak{u}(1)\oplus\mathbb{C}^{0|32}\cong\mathbb{C}^{31|32}.
\end{equation}

Consider the super-conformal algebra $\mathfrak{psu}(2,2|4)$, which is just the real realization of this $\mathfrak{psl}(4|4)$ algebra, acting on the twistor space $\mathbb{CP}^{3|4}$, which will act as first order differential operators that are invariant under re-scaling. Denoting the coordinates as $(\lambda^\alpha;\mu^{\dot\alpha},\psi^A)$, the bosonic generators are:
\begin{align}
&p^{\alpha}_{\dot\alpha} = i\lambda^\alpha\frac{\partial}{\partial\mu^{\dot\alpha}}\hspace{1cm}\mbox{(Translation)}\label{t}\\
&k^{\dot\alpha}_{\alpha}=i\mu^{\dot\alpha}\frac{\partial}{\partial\lambda^\alpha}\hspace{1cm}\mbox{(SCT)}\label{sct}\\
j^\alpha_\beta = \frac{i}{2}\left( \lambda_\alpha\frac{\partial}{\partial\lambda^\beta} + \lambda_\beta\frac{\partial}{\partial\lambda^\alpha}\right)&,\ \ \ j^{\dot\alpha}_{\dot\beta} = \frac{i}{2}\left( \mu_{\dot\alpha}\frac{\partial}{\partial\mu^{\dot\beta}}+ \mu_{\dot\beta}\frac{\partial}{\partial\mu^{\dot\alpha}}\right)\hspace{1cm}\mbox{(Rotations)}\label{rot}\\
& d = \frac{i}{2}\left( \lambda^\alpha\frac{\partial}{\partial\lambda^\alpha} + \mu^{\dot\alpha}\frac{\partial}{\partial\mu^{\dot\alpha}}\right)\hspace{1cm}\mbox{(Dilation)}\label{d}
\end{align}
while the fermionic part is generated by the following super-charges:
\begin{align}
\hspace{1cm}&s^A_\alpha=i\psi^A\frac{\partial}{\partial\lambda^{\alpha}},\ \ \ \ \bar s_A^{\dot\alpha}=\mu^{\dot\alpha}\frac{\partial}{\partial\psi^A},\hspace{1cm}\mbox{(Super-conformal)}\\
&q^\alpha_A = i\lambda^{\alpha}\frac{\partial}{\partial\psi^A},\ \ \ \ \bar q^A_{\dot\alpha} =\psi^A\frac{\partial}{\partial\mu^{\dot\alpha}}\hspace{1.2cm}\mbox{(Super-symmetry)}
\end{align}
The $SU(4)$ R-symmetry is:
\begin{equation}
r^A_B = \psi^A\frac{\partial}{\partial \psi^B} - \frac{1}{4}\delta^A_B\psi^C\frac{\partial}{\partial \psi^C}
\end{equation}

\subsection{PSL algebra}

The super-conformal algebra $\mathfrak{psu}(2,2|4)$ can be thought of as the real form of $\mathfrak{psl}(4|4)$ in Lorentz signature. By itself, this algebra can be represented as linear operators in $\mathbb{C}^{4|4}$, as explained above. In this representation, we may find the structure constant explicitly. First, we parametrize $\mathbb{C}^{4|4}$ by $\Phi^\mathcal{I} = \{ Z^I,\psi^A\}$:

\begin{equation}
\left[Z^I\frac{\partial}{\partial Z^K},Z^J\frac{\partial}{\partial Z^L}\right] = \delta^J_KZ^I\frac{\partial}{\partial Z^L} - \delta^I_LZ^J\frac{\partial}{\partial Z^K} = f_{^{IJ}_{KL}}^{\ \ \ ^M_N}Z^M\frac{\partial}{\partial Z^N} \nonumber
\end{equation}
\begin{equation}
\left[\psi^A\frac{\partial}{\partial \psi^B},\psi^C\frac{\partial}{\partial \psi^D}\right] = \delta^C_B\psi^A\frac{\partial}{\partial \psi^D} - \delta^A_D\psi^C\frac{\partial}{\partial \psi^B} = f_{^{AC}_{BD}}^{\ \ \ ^E_F}\psi^E\frac{\partial}{\partial \psi^F} \nonumber
\end{equation}
\begin{equation}
\left[\psi^A\frac{\partial}{\partial Z^I},Z^J\frac{\partial}{\partial Z^L}\right] = \delta^J_I\psi^A\frac{\partial}{\partial Z^L} = f_{^{AJ}_{IL}}^{\ \ \ ^B_K}\psi^B\frac{\partial}{\partial Z^K} \nonumber
\end{equation}
\begin{equation}
\left[Z^I\frac{\partial}{\partial \psi^A},Z^J\frac{\partial}{\partial Z^L}\right] =  - \delta^I_LZ^J\frac{\partial}{\partial \psi^A} = f_{^{IJ}_{AL}}^{\ \ \ ^K_B}Z^K\frac{\partial}{\partial \psi^B}\nonumber
\end{equation}
\begin{equation}
\left\{\psi^A\frac{\partial}{\partial Z^J},Z^I\frac{\partial}{\partial \psi^B}\right\} = \delta^I_J\psi^A\frac{\partial}{\partial \psi^B} + \delta^A_BZ^I\frac{\partial}{\partial Z^J} = f_{^{AI}_{JB}}^{\ \ \ ^C_D} \psi^C\frac{\partial}{\partial \psi^D} + f_{^{AI}_{JB}}^{\ \ \ ^K_L}Z^K\frac{\partial}{\partial Z^L}\nonumber
\end{equation}
where we have also a vanishing part:
\begin{equation}
\left[Z^I\frac{\partial}{\partial Z^K},\psi^A\frac{\partial}{\partial \psi^B}\right] = \left\{\psi^A\frac{\partial}{\partial Z^K},\psi^B\frac{\partial}{\partial Z^L}\right\} = \left\{ Z^I\frac{\partial}{\partial \psi^A},Z^J\frac{\partial}{\partial \psi^B} \right\} = 0\nonumber
\end{equation}
in other words, the non-vanishing structure constants are, for the even part of the algebra:
\begin{align}
f_{^{IK}_{JL}}^{\ \ \ ^M_N} = \delta^K_J\delta^I_M\delta^N_L-\delta_L^I\delta^K_M\delta^N_J&,\ \ \ \ f_{^{AC}_{BD}}^{\ \ \ ^E_F} = \delta^C_B\delta^A_E\delta^F_D-\delta_D^A\delta^C_E\delta^F_B, \\
f_{^{AJ}_{IL}}^{\ \ \ ^B_K} = \delta^J_I\delta^A_B\delta^K_L&,\ \ \ \ f_{^{JI}_{LA}}^{\ \ \ ^K_B} = \delta^I_L\delta^J_K\delta^B_A,
\end{align}
and for the odd part of the algebra:
\begin{align}
f_{^{AI}_{JB}}^{\ \ \ ^C_D} =  \delta^I_J\delta^A_C\delta^D_B&, \ \ \ f_{^{AI}_{JB}}^{\ \ \ ^K_L} =  \delta^A_B\delta^I_K\delta_J^L,
\end{align}
summarizing:
\begin{equation}\label{f}
f_{^{\mathcal{I}\mathcal{K}}_{\mathcal{J}\mathcal{L}}}^{\ \ \ ^\mathcal{M}_\mathcal{N}} = \delta^\mathcal{K}_\mathcal{J}\delta^\mathcal{I}_\mathcal{M}\delta^\mathcal{N}_\mathcal{L}- (-1)^{\left|^\mathcal{I}_\mathcal{J}\right|\cdot\left|^\mathcal{K}_\mathcal{L}\right|}\delta_\mathcal{L}^\mathcal{I}\delta^\mathcal{K}_\mathcal{M}\delta^\mathcal{N}_\mathcal{J}
\end{equation}
where $\left|^I_J\right|=\left|^A_B\right|=0$ and $\left|^I_A\right|=\left|^A_I\right|=1$.

One may realize the $\frak{psl}$ algebra in a more familiar way by going to the local coordinates. First, we set the linear operators described above in a block form matrix. For convenience, we replace the notation of the bosonic coordinates: $Z^I\to\phi^I$, and then write the matrix:
\begin{equation}
\Phi^{\cal I}\frac{\partial}{\partial\Phi^{\cal J}} = \begin{pmatrix}
\ & \phi^I\frac{\partial}{\partial\phi^J}& \ & \ &\psi^A\frac{\partial}{\partial\phi^J} & \\
\ & \ & \ & \ & \ & \\
\ & \phi^I\frac{\partial}{\partial\psi^B}& \ & \ &\psi^A\frac{\partial}{\partial\psi^B} &
\end{pmatrix}
\end{equation}
We can identify the elements of the super-conformal algebra in the block matrix:
\begin{equation}\label{conformalmatrix}
\begin{matrix}
\includegraphics[width=0.5\linewidth]{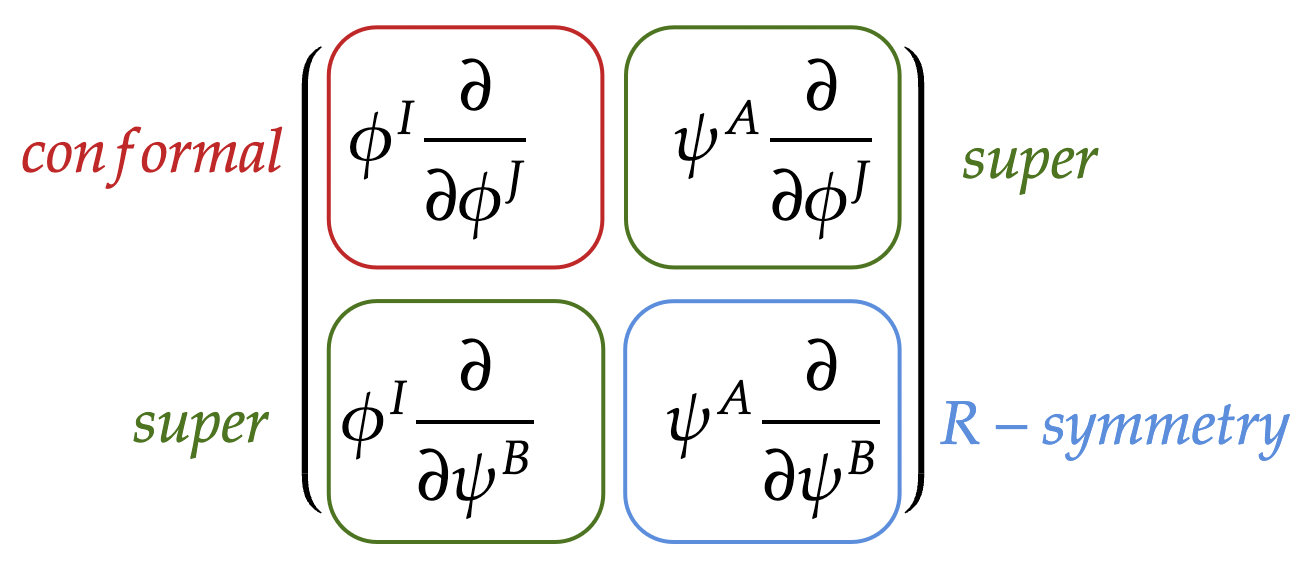}
\end{matrix}
\end{equation}

\section{Cohomology computation}

In order to give a more rigorous derivation of the cohomology studied in section 4, we will compute the sheaf cohomology directly, using the Euler short exact sequence and its corresponding long exact sequence.

\subsection{Ghost number 1 operators}

To study the ghost number 1 operators, we use the Euler short exact sequence \cite{CPn},\cite{Noja}:
\begin{equation}\label{euler}
0\longrightarrow \mathcal{O}_{\mathbb{CP}^{3|4}} \longrightarrow \mathcal{O}_{\mathbb{CP}^{3|4}}(1)\otimes\mathbb{C}^{4|4}\longrightarrow T{\mathbb{CP}^{3|4}} \longrightarrow 0
\end{equation}
The space $\mathcal{O}_{\mathbb{CP}^{3|4}}(1)$ is defined as the set of homogeneous functions $\ell_i$, defined on $\mathbb{C}^{4|4}$, with degree $1$, i.e., functions satisfying the condition $\ell_i(tZ) = t\ell_i(Z)$, for any complex number $t$, where $Z^i$ are the coordinates of $\mathbb{C}^{4|4}$. Then, the tensor product $\mathcal{O}_{\mathbb{CP}^{3|4}}(1)\otimes\mathbb{C}^{4|4}$ contains vectors of the form $(\ell_1,\cdots,\ell_8)$. The firs arrow of this sequence is an injective map defined by:
\begin{align}
\begin{split}
&\mathcal{O}_{\mathbb{CP}^{3|4}} \longrightarrow \mathcal{O}_{\mathbb{CP}^{3|4}}(1)\otimes\mathbb{C}^{4|4}\hspace{1.2cm}\\
& c\longmapsto c\cdot(Z_1,\cdots,Z_8),
\end{split}
\end{align}
The second map is a surjective function, given by:
\begin{align}\label{map2}
\begin{split}
\hspace{1cm}&\mathcal{O}_{\mathbb{CP}^{3|4}}(1)\otimes\mathbb{C}^{4|4}\longrightarrow T{\mathbb{CP}^{3|4}}\\
(\ell_1&(Z),\cdots,\ell_8(Z))\longmapsto \ell_1(Z)\frac{\partial}{\partial Z^1}+\cdots+\ell_8(Z)\frac{\partial}{\partial Z^8}
\end{split}
\end{align}

What this short exact sequence says is that the tangent bundle is given by equivalent classes of vector fields in $\mathbb{C}^{4|4}$. Since the second map is surjective, the vector fields in the tangent bundle are all vector fields with coefficients given by homogeneous functions, as described in the last map (\ref{map2}). To figure out what is the kernel of this surjective map, one notice that this sequence splits and therefore the kernel comes from the first set of the sequence. The first set of the sequence is mapped to the Euler vector field $E=Z^i\frac{\partial}{\partial Z^i}$. Therefore, one excludes the Euler vector field from the tangent bundle by considering it zero in the equivalence relation: $E\sim 0$.

We can consider now the corresponding cohomology long exact sequence. Since $H^1(\mathcal{O}_{\mathbb{CP}^{3|4}})=0$ \cite{Noja}, the first part of the sequence is:
\begin{equation}
0\longrightarrow H^0(\mathbb{CP}^{3|4},\mathcal{O}_{\mathbb{CP}^{3|4}}) \longrightarrow H^0\left(\mathbb{CP}^{3|4},\mathcal{O}_{\mathbb{CP}^{3|4}}(1)\otimes\mathbb{C}^{4|4}\right)\longrightarrow H^0(\mathbb{CP}^{3|4},T{\mathbb{CP}^{3|4}}) \longrightarrow 0
\end{equation}
from which follows that
\begin{align}\label{H0T}
\begin{split}
H^0(\mathbb{CP}^{3|4},T{\mathbb{CP}^{3|4}}) &\cong \frac{H^0(\mathbb{CP}^{3|4},\mathcal{O}_{\mathbb{CP}^{3|4}}(1)\otimes \mathbb{C}^{4|4})}{H^0(\mathbb{CP}^{3|4},\mathcal{O}_{\mathbb{CP}^{3|4}})}
\end{split}
\end{align}
The denominator $H^0(\mathbb{CP}^{3|4},\mathcal{O}_{\mathbb{CP}^{3|4}})$ consists of the Euler vector field: $E=Z^I\frac{\partial}{\partial Z^I}$. The numerator is $H^0\Big(\mathbb{CP}^{3|4},\mathcal{O}_{\mathbb{CP}^{3|4}}(1)\Big)\otimes \mathbb{C}^{4|4}$ and consists of the set of homogeneous global functions in the projective space of degree 1, generated by $\{Z^i\}_{i=1,\cdots,8}$, tensored by $\mathbb{C}^{4|4}$. Therefore, this set is generated by vectors of the form $(Z^{i_1},\cdots,Z^{i_8})$, which under the second arrow in the Euler exact sequence (\ref{euler}) is mapped to a vector field of the form $Z^{i_1}\frac{\partial}{\partial Z^1}+\cdots+Z^{i_8}\frac{\partial}{\partial Z^8}$. In conclusion, the numerator in (\ref{H0T}) is give by elements of $\mathfrak{gl}(4|4)$, and the denominator is $\mathfrak{u}(1)$, therefore:
\begin{equation}
H^0(\mathbb{CP}^{3|4},T{\mathbb{CP}^{3|4}}) \cong \frac{\mathfrak{gl}(4|4)}{\mathfrak{u}(1)}=\mathfrak{pgl}(4|4)
\end{equation}

\subsection{BRST cohomology of ghost number 2}

For the second exterior power, we consider, the dual of the Euler sequence (\ref{euler}):
\begin{equation}
\begin{tikzcd}
0\arrow{r}{}&\Omega^1({\mathbb{CP}^{3|4}})\arrow[r, "\iota"]&\mathcal{O}_{\mathbb{CP}^3}(-1)^{4|4}\arrow[r, "\pi"]& \mathcal{O}_{\mathbb{CP}^3}\arrow{r}{}&0,
\end{tikzcd}
\end{equation}
and, as was done in \cite{Noja}, we have the following corresponding exact sequence:
\begin{equation}\label{wedgedual}
\begin{tikzcd}
0\arrow{r}{}&\bigwedge\nolimits^2 \Omega^1_{\mathbb{CP}^{3|4}}\arrow[r, "\wedge^2\iota"]&\bigwedge\nolimits^2\mathcal{O}_{\mathbb{CP}^{3|4}}(-1)^{4|4}\arrow[r, "\phi"]& \Omega^1_{\mathbb{CP}^{3|4}}\arrow{r}{}&0
\end{tikzcd}
\end{equation}
where the map $\phi$ is constructed as follows: since the original sequence splits, we have an isomorphism $\mathcal{O}_{\mathbb{CP}^{3|4}}(-1)^{4|4}\cong \Omega^1({\mathbb{CP}^{3|4}})\oplus \mathcal{O}_{\mathbb{CP}^{3|4}}$. Taking the exterior power, one gets:
\begin{equation}
\bigwedge\nolimits^2 \mathcal{O}_{\mathbb{CP}^{3|4}}(-1)^{4|4}\cong \bigwedge\nolimits^2\Omega^1_{\mathbb{CP}^{3|4}} \oplus (\Omega^1_{\mathbb{CP}^{3|4}}\otimes \mathcal{O}_{\mathbb{CP}^{3|4}})\oplus \bigwedge\nolimits^2 \mathcal{O}_{\mathbb{CP}^{3|4}} \cong \Big(\bigwedge\nolimits^2\Omega^1_{\mathbb{CP}^{3|4}}\Big) \oplus \Omega^1_{\mathbb{CP}^{3|4}}
\end{equation}
and therefore we have:
\begin{equation}
\bigwedge\nolimits^2 \mathcal{O}_{\mathbb{CP}^{3|4}}(-1)^{4|4}\Big/ \bigwedge\nolimits^2\Omega^1_{\mathbb{CP}^{3|4}} \cong \Omega^1_{\mathbb{CP}^{3|4}}.
\end{equation}

We now take the dual of this sequence $(\ref{wedgedual})$:

\begin{equation}
0\longrightarrow T{\mathbb{CP}^{3|4}}\longrightarrow \bigwedge\nolimits^2\mathcal{O}_{\mathbb{CP}^{3|4}}(1)^{\oplus4|4} \longrightarrow \bigwedge\nolimits^2T{\mathbb{CP}^{3|4}} \longrightarrow 0
\end{equation}
and from its corresponding cohomology long exact sequence one has, using $H^1\left(T{\mathbb{CP}^{3|4}}\right)=0$, that:
\begin{equation}\label{sequence^2}
0\longrightarrow H^0\left(\mathbb{CP}^{3|4},T{\mathbb{CP}^{3|4}}\right)\longrightarrow H^0\left(\mathbb{CP}^{3|4},\bigwedge\nolimits^2\mathcal{O}_{\mathbb{CP}^3}(1)^{\oplus4|4}\right) \longrightarrow H^0\left(\mathbb{CP}^{3|4},\bigwedge\nolimits^2T{\mathbb{CP}^{3|4}}\right)\longrightarrow 0
\end{equation}

and finally we have the isomorphism:
\begin{equation}\label{H0^2}
H^0\left(\mathbb{CP}^{3|4},\bigwedge\nolimits^2T{\mathbb{CP}^{3|4}}\right) \cong \frac{H^0\left(\mathbb{CP}^{3|4},\bigwedge\nolimits^2\mathcal{O}_{\mathbb{CP}^{3|4}}(1)^{\oplus4|4}\right)}{H^0\left(\mathbb{CP}^{3|4},T{\mathbb{CP}^{3|4}}\right)}
\end{equation}

The numerator is a set given by the antisymmetric product of two vectors of 1-homogeneous functions: $\Vec{\ell}(tZ) = t\Vec{\ell}(Z)$. The basis to these functions are $Z^I$, for $I=1,\cdots,8$:
\begin{equation}
\bigwedge\nolimits^2\mathcal{O}_{\mathbb{CP}^{3|4}}(1)^{\oplus 4|4} = \mathcal{O}_{\mathbb{CP}^{3|4}}(2)\otimes \bigwedge\nolimits^2\mathbb{C}^{4|4}
\end{equation}
and therefore:
\begin{equation}
H^0\left(\mathbb{CP}^{3|4},\bigwedge\nolimits^2\mathcal{O}_{\mathbb{CP}^{3|4}}(1)^{\oplus 4|4}\right) = H^0\Big(\mathbb{CP}^{3|4},\mathcal{O}_{\mathbb{CP}^{3|4}}(2)\Big)\otimes \bigwedge\nolimits^2\mathbb{C}^{4|4}
\end{equation}
The global holomorphic sections $\mathcal{O}_{\mathbb{CP}^{3|4}}(2)$ correspond to homogeneous polynomials in two variables:
\begin{equation}
H^0\Big(\mathbb{CP}^{3|4},\mathcal{O}_{\mathbb{CP}^{3|4}}(2)\Big) \cong \Big\{ Z^IZ^J\ |\ I,J=1,\cdots,8 \Big\}
\end{equation}

Therefore, the vector fields representing this set are:
\begin{align}\label{zdz^zdz}
Z^IZ^J\frac{\partial}{\partial Z^K} \wedge \frac{\partial}{\partial Z^L}
\end{align}

Since this can be written as the product of two elements of the basis of the $\mathfrak{gl}$-algebra, $t^I_J=Z^I\frac{\partial}{\partial Z^J}$, the denominator is the antisymmetric product $\mathfrak{gl}(4|4)\wedge \mathfrak{gl}(4|4)$. Let us now study the denominator in (\ref{H0^2}) in order to properly describe the entire cohomology group. Denoting by $\{ e_I\wedge e_J\}_{I<J}$ the basis of $\mathbb{C}^{4|4}$, the first map of the exact sequence (\ref{sequence^2}) is:
\begin{equation}
[t^I_J]=Z^I\frac{\partial}{\partial Z^J} \mapsto Z^IZ^P\otimes (e_J \wedge e_P)
\end{equation}
where I took a generic representation $t^I_J\in H^0\left(\mathbb{CP}^{3|4},T{\mathbb{CP}^{3|4}}\right)\cong\mathfrak{pgl}(4|4)$. The second map is the following:
\begin{equation}
Z^IZ^K\otimes (e_J\wedge e_L) \mapsto  Z^IZ^K\frac{\partial}{\partial Z^J} \wedge \frac{\partial}{\partial Z^L}
\end{equation}
and therefore the denominator is the map of $t^I_J$ under these 2 arrows:
\begin{equation}
Z^IZ^P\frac{\partial}{\partial Z^J} \wedge \frac{\partial}{\partial Z^P}
\end{equation}
In conclusion, we factor out the elements of the form above. This means that not all elements of the form (\ref{zdz^zdz}) are in the cohomology, and we have to exclude two types of then:
\begin{align}
\begin{split}
&Z^I\frac{\partial}{\partial Z^J}\wedge Z^P\frac{\partial}{\partial Z^P}\ \ \ \text{and}\ \ \  Z^I\frac{\partial}{\partial Z^P}\wedge Z^P\frac{\partial}{\partial Z^J}.
\end{split}
\end{align}

The first one is excluded by replacing $\mathfrak{gl}\wedge\mathfrak{gl}$ by $\mathfrak{pgl}\wedge\mathfrak{pgl}$ in the denominator of our cohomology group. The second one in excluding the elements of the form:
\begin{equation}
f_{^I_J\ ^K_L}^{\ \ \ \ ^M_N}Z^I\frac{\partial}{\partial Z^J}\wedge Z^K\frac{\partial}{\partial Z^L} = (\delta^I_L\delta^M_J\delta^K_N-\delta_J^K\delta^M_L\delta^I_N)Z^I\frac{\partial}{\partial Z^J}\wedge Z^K\frac{\partial}{\partial Z^L} = 2Z^I\frac{\partial}{\partial Z^M}\wedge Z^N\frac{\partial}{\partial Z^I}\nonumber
\end{equation}

Therefore, we finally have a characterization for our cohomology:

\begin{equation}
H^0\left(\mathbb{CP}^{3|4},\wedge^2T{\mathbb{CP}^{3|4}}\right) \cong\frac{\mathfrak{pgl}(4|4)\wedge\mathfrak{pgl}(4|4)}{\mathfrak{pgl}(4|4)}
\end{equation}
where the quotient in this algebraic description means to divide by elements of the form $f_{bc}^{\ \ a}t^b\wedge t^c$, where $t_a\in\mathfrak{g}$.

\providecommand{\href}[2]{#2}\begingroup\raggedright\endgroup


\end{document}